\documentclass[prb,aps,twocolumn,longbibliography, nofootinbib, showpacs]{revtex4-2}
\usepackage[colorlinks=true, citecolor=blue, urlcolor=blue, linkcolor = blue ]{hyperref}
\usepackage{epsfig,newlfont,amssymb,amsfonts,amsmath,bm,subfigure,palatino,mathtools,amsthm,braket,times,soul,enumitem,color}
\usepackage[normalem]{ulem}
\newcommand{\stkout}[1]{\ifmmode\text{\sout{\ensuremath{#1}}}\else\sout{#1}\fi}
\usepackage[T1]{fontenc}
\usepackage[normalem]{ulem}
\usepackage{graphicx}
\usepackage{placeins}
\usepackage{hyperref}
\usepackage{multirow}
\usepackage{hhline,booktabs}
\usepackage{xcolor}
\usepackage{bbm}
\usepackage{comment}
\usepackage{booktabs}
\usepackage{multirow}
 \usepackage{array}

\DeclareMathAlphabet{\mathpzc}{OT1}{pzc}{m}{it}

\begin{document}

\preprint{}

\title{Entanglement asymmetry and quantum Mpemba effect in string-dressed long-range spin chains}

\author{Debarshi Paul$^{1,2,3}$,
Debashish Mallick$^{1,2,4}$,
Soumyadip Mukherjee$^{1,2,5}$,
Ayan Sahoo$^{1,2}$,
Debraj Rakshit$^{1,2}$}

\affiliation{$^1$Harish-Chandra Research Institute,
Allahabad 211019, India}

\affiliation{$^2$Homi Bhabha National Institute,
Training School Complex, Anushakti Nagar,
Mumbai 400094, India}

\affiliation{$^3$ Department of Physical Sciences, Indian Institute of Science Education and Research Kolkata, Mohanpur, Nadia-741246, India}

\affiliation{$^4$Institute of Mathematical Sciences,
CIT Campus, Chennai 600113, India}

\affiliation{$^5$ Quantum Research Park (QuRP), FSID, Indian Institute of Science, Bengaluru 560012, India}

\begin{abstract}
The quantum Mpemba effect has been studied in short-range integrable systems and separately, in the long-range non-integrable systems. We bridge the two ends of the interaction range with a string-dressed XY spin chain of tunable range that is exactly solvable throughout and track how interaction range reshapes late-time dynamics. In equilibrium, the entanglement asymmetry of the ground state grows logarithmically with the subsystem size in all interaction regimes, and it acts as an indicator of the quantum phase transition. We consider two classes of initial states with broken $U(1)$-symmetry, the tilted ferromagnetic states and the ground state of the anisotropic spin chain, and study dynamical restoration of the symmetry after performing a sudden quench with the symmetry-preserving spin chain.
For the tilted ferromagnetic states, pairs whose tilt angles sum to more than the straight angle display quantum Mpemba effect in all interaction ranges, whereas a pair with a smaller sum may show  quantum Mpemba effect only in the short-range. For the ground state of the anisotropic chain, it occurs for every interaction range. The results for entanglement asymmetry are interpreted within the quasiparticle picture, and a microscopic understanding is followed from the late-time asymmetry carried by the slowest quasiparticle pairs, a mechanism distinct from that of the bare long-range chain. The interaction range decides which pairs are slow. They reside near the zone boundary for every interaction range, but an additional contribution arises near zero momentum above a threshold value of the power-law decay exponent, corresponding to the onset of the short-range regime. We derive closed-form asymptotic expressions for the entanglement asymmetry at late times, separately below and above the onset of the short-range regime, and provide explicit criteria for the occurrence and absence of the quantum Mpemba effect for both families of states.

\end{abstract}

\maketitle
 \section{Introduction}  

The Mpemba effect is an anomalous relaxation phenomenon in which a system prepared farther from equilibrium relaxes faster toward equilibrium than the system when prepared closer to equilibrium. Since its modern formulation as an anomalous relaxation phenomenon in classical systems in 1969 \cite{mpemba1969cool}, the Mpemba effect has been experimentally observed in several classical platforms \cite{Auerbach1995, Jeng2006, Burridge2016, Kumar2020, Kumar2022}  and theoretically explained via a variety of nonequilibrium relaxation frameworks~\cite{PhysRevLett.119.148001, Lu2017, PhysRevX.9.021060, PhysRevE.102.012906, Biswas2021, Holtzman2022, Santos2020}.

 Its quantum extension appeared only recently with studies focusing on various aspects of open quantum dynamics \cite{PhysRevLett.127.060401,PhysRevLett.131.080402,PhysRevA.110.022213, PhysRevResearch.6.033330,PhysRevLett.133.136302,PhysRevA.110.042218, 8fp4-15mq, das2025rolereversalquantummpemba, bagui2025detectionmpembaeffectgood} and closed many-body systems, where relaxation is not driven by external dissipation but emerges from unitary many-body dynamics \cite{ares2023entanglement, Murciano2024,PhysRevLett.133.010401,Lastres_2025, PhysRevB.110.085126, Chalas_2024, PhysRevB.111.104312, PhysRevA.111.043304,l4xx-plmv, 1td3-2vwf,fujimura2026entanglementasymmetryquantummpemba, yu2025quantum,l4xx-plmv}. In this context, the eigenstate thermalization hypothesis (ETH)  provides the conceptual basis for understanding how local observables and reduced subsystems can approach stationary, thermal-like states even though the global state evolves unitarily \cite{alishahiha2025eigenstatethermalizationhypothesisshort, Deutsch_2018}, thereby allowing Mpemba-like anomalous relaxation to be diagnosed in isolated quantum systems. 

 The Mpemba effect is characterized through a variety of quantum measures quantifying relaxation, e.g., distance measures from the stationary state \cite{PhysRevLett.127.060401, PhysRevLett.133.140404, PhysRevLett.133.136302}. A particularly useful formulation of the Quantum Mpemba Effect (QME) in closed systems is based on symmetry restoration \cite{ares2023entanglement, Murciano2024, PhysRevLett.133.010401}. In this setting, the system is initialized in states that break a conserved symmetry of the post-quench Hamiltonian.

 The QME is then identified when an initially more symmetry-broken state restores the symmetry faster than a less symmetry-broken one. 

In this formulation, the role of the distance from the symmetry-restored stationary regime is played by the Entanglement Asymmetry (EA), a subsystem measure of symmetry breaking defined by the mismatch between a reduced density matrix and its symmetry-projected counterpart \cite{ares2023entanglement}. This mechanism has been analyzed in spin chains, free-fermion systems, random circuits, and higher-dimensional fermionic models, establishing EA as a natural diagnostic of symmetry-based anomalous relaxation in closed quantum many-body systems \cite{Ares2025}.

{\color{black}Experimentally, open-system QME have been observed with trapped ions~\cite{AharonyShapira2024,zhang2025observation} and in few-qubit
platforms~\cite{schnepper2025experimentalobservationapplicationgenuine,chatterjee2025directexperimentalobservationquantum}. The symmetry-restoration QME has been demonstrated in a trapped-ion quantum simulator with tilted ferromagnetic states as initial states and XX-chain with power-law couplings as the Hamiltonian governing the quenched dynamics \cite{PhysRevLett.133.010402}. Subsequently, tunable-range couplings in superconducting processors have demonstrated the
QME phenomenon when the nearest-neighbour couplings dominate, while its suppression has been reported when short- and long-range couplings are comparable \cite{Xu2026}. Rydberg arrays, polar molecules, and dipolar gases~\cite{RevModPhys.95.035002, RevModPhys.82.2313, Lahaye_2009, Carr_2009, Blatt:2012chk, 2008Natur.453.1008B, RevModPhys.93.025001} are naturally characterized by long-range interactions and provide essential ingredients for probing long-range effects on the QME.

The role of the interaction range has been investigated theoretically. In
$U(1)$-symmetric long-range spin chains, tilted ferromagnetic states display the QME generically \cite{Yamashika2026}. This has been explained by employing time-dependent spin-wave theory. There, the phenomenon is driven by zero-momentum spin waves, and the underlying mechanism is distinct from the quasiparticle transport. In the nonintegrable long-range XYZ chains, the QME in the emergent $U(1)$-symmetry in the prethermal regime is found in sufficiently short-range interactions as a consequence of the Hohenberg--Mermin--Wagner theorem~\cite{hallam2025tunablequantummpembaeffect}. Tunable QME has been reported in long-range $U(1)$-symmetric random circuits~\cite{l4xx-plmv}. Ground states of long-range Kitaev chains quenched to short-range free fermions exhibit multiple crossings of the asymmetry, so that the QME can only be inferred from the late-time behaviour~\cite{Chalas_2024}. In these settings, the power law either breaks integrability or enters via the initial state, and the quasiparticle mechanism that underlies the short-range QME is not available to track how the interaction range reshapes the late-time dynamics. In this work, we close that gap with a long-range XY chain whose couplings are dressed with Jordan-Wigner strings.

The string dressing preserves the quadratic Jordan-Wigner structure, making the model exactly solvable,  while the power-law couplings introduce a tunable long-range effect. Thus, this setting provides a controlled bridge between the short-range integrable formulation of QME and more general long-range many-body relaxation. Based on the small-momentum dispersion behaviour, the power-law exponent $\alpha$ controls the interaction profile and distinguishes three regimes, where the system is long-ranged or is effectively short- or quasi-long-ranged. It also controls the criticality in the system: the critical field remains finite for $\alpha>1$; However,  for $\alpha<1$, it grows with an interaction cut-off limiting the range of interaction, but becomes infinite when the cut-off is removed.

We first perform a static analysis of the ground-state EA, which signals the quantum phase transition through a kink at the critical field,  for every interaction range. The same
analysis is later used to order initial asymmetries entering the QME for the class of ground states as the initial states. Apart from the ground states, we investigate the tilted ferromagnetic states as another class of initial states. We show that the QME phenomena can be reshaped by tuning the interaction range. Moreover, this allows the quasi-particle interpretation of symmetry-restoration-based QME for every $\alpha$. We consider tilted ferromagnetic states and ground states of the anisotropic chain as the initial state, and quench to the $U(1)$-symmetric long-range XX chain and monitor the EA. For the tilted ferromagnetic class of states, we show that if the sum of the tilt angles of two initial states is below $\pi$, the QME is absent in the long-range regime, and it appears only in the short-range, for suitable pairs. If the angles sum to more than $\pi$, the QME occurs for every interaction range.
 
For ground states of the anisotropic chain a suitable pairs of initial states display the QME in all interaction ranges. Thus, the occurrence of the QME depends on the structure of the initial state as well as on the interaction range.  Using the quasiparticle picture, we provide a closed-form expression for the late-time asymptotic expressions of the asymmetry in every range. }

The rest of the paper is organized as follows. Section II introduces the model and analytical framework. Section III examines the effects of long-range interactions on the ground-state EA. Section IV presents the QME for tilted and ground states, while Sec. V explains the QME in terms of quasiparticle density and velocity. Finally, Sec. VI presents the discussion and conclusions.
\begin{figure}[t]
    \centering
    \includegraphics[width=\columnwidth]{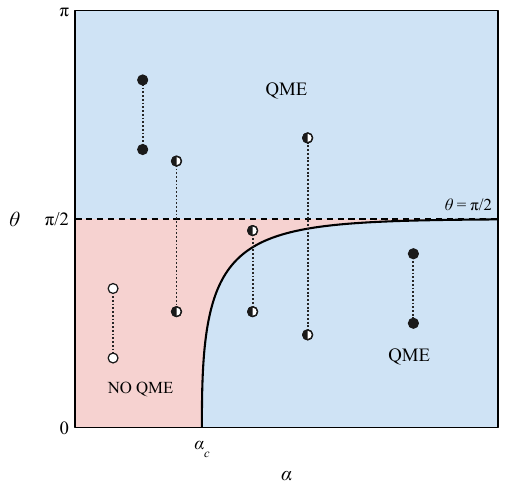}
    \caption{The schematic diagram illustrates the QME for the tilted ferromagnetic state with tilt angle $\theta\in(0,\pi)$. Each vertical dotted line joins a pair of initial states parametrised by $\theta$. The open circles denote pairs of states not showing QME and the filled circles denote pairs showing QME. The half-filled circles are for pairs for which the QME may or may not occur depending on the specific values of the tilt-angle $\theta$. Above the dashed line, $\theta=\pi/2$, every pair of states displays QME for any value of interaction range parameter $\alpha$. Below, $\theta=\pi/2$, QME requires $\alpha>\alpha_c$ and occurs for pairs $(\theta_A,\theta_B)$ below a threshold $\theta^*(\alpha)$(solid line). For $\alpha<\alpha_c$, given a pair $\theta_A<\theta_B<\pi/2$ , there is no QME.}
    \label{fig: QME schematic}
\end{figure}

\section{Model}

The one-dimensional long-range XY spin chain  of length $N$ with algebraically decaying interactions  $J_r \propto r^{-\alpha}$ is described by the Hamiltonian as \cite{Vodola_2016},

{\color{black}
\begin{equation}
\begin{aligned}
H = & - \sum_{j=1}^{N} \sum_{r=1}^{N_c} J_r
\Bigg[
\frac{1+\gamma}{4}\,\sigma_j^x \mathcal{Z}_r^z \sigma_{j+r}^x \\
& \qquad\qquad + \frac{1-\gamma}{4}\,\sigma_j^y \mathcal{Z}_r^z \sigma_{j+r}^y
\Bigg]
- \frac{h}{2} \sum_{j=1}^{N} \sigma_j^z,
\end{aligned}
\label{Hamiltonian}
\end{equation}
where $N_c \le N/2$}, $J$ is the interaction strength, $\alpha$ controls the power-law decay of interactions, $\gamma$ is the anisotropy parameter, and $h$ denotes the transverse field strength. The operators $\sigma_j^{x,y,z}$ are Pauli spin matrices, while $\mathcal{Z}_r^z=\prod_{k=j+1}^{j+r-1}\sigma_k^z$ is the Jordan--Wigner string operator. Here, $J_r = J/r^\alpha$, where $\alpha$ denotes the power-law decay exponent of the interaction. The parameter $\gamma$ represents the XY anisotropy, and $h$ is the external magnetic field.   {\color{black} Throughout this paper, we have used $N_c=N/2$, with the exact value $N_c=2\times10^6$ for numerical calculations. It is chosen such that the subsystem sizes, $l$, that we probe later in this work, remains much smaller than $N_c$, i.e., $l\ll N_c$.}

The exponent $\alpha$ is commonly used to distinguish strong long-range $\alpha < 1$, weak long-range, and effectively short-range interactions \cite{RevModPhys.95.035002,Fraxanet2021,Konar2025}. In the limit $\alpha \to \infty$, the model reduces to the nearest-neighbour model. For the one-dimensional kitaev chain, the regimes of $\alpha>2$ and $1<\alpha<2$ are differentiated by the decay of the correlations \cite{Vodola_2016}.  For non-interacting particles with power-law interaction, the threshold of $\alpha>2$ marks the appearance of a linear light cone for free particles \cite{Tran2020}. For interacting systems, a linear Lieb–Robinson light cone requires $\alpha>3$ in one dimension \cite{Tran2020}. In the present model, as we discuss below, the dynamics is controlled by the dispersion. The dispersion is analytic, and the group velocity is linear at $k = 0$ for $\alpha>3$. We therefore refer to $\alpha<1$ as long-range, $1<\alpha<3$ as quasi-long-range, and $\alpha>3$ as short-range, with the understanding that the boundary $\alpha=3$ is set by the analyticity of the dispersion and linearity of the group velocity.

Under periodic boundary conditions, this long-range model can be solved analytically by mapping the spin operators to fermionic operators via the Jordan–Wigner transformation: 
{\color{black} 
\begin{equation}
\begin{aligned}
\sigma_j^z &= 1 - 2c_j^\dagger c_j, \\
\sigma_j^x &= \left(\prod_{k<j}(1-2n_k)\right)(c_j^\dagger + c_j), \\
\sigma_j^y &= i\left(\prod_{k<j}(1-2n_k)\right)(c_j^\dagger - c_j).
\end{aligned}
\end{equation}
}

In the fermionic representation, the Hamiltonian takes the form
{\color{black}
\begin{equation}
\begin{aligned}
    H &= -\frac{1}{2}\sum_{j=1}^{N}\sum_{r=1}^{N_c} J_r \left[c_j^\dagger c_{j+r} + \gamma c_j^\dagger c_{j+r}^\dagger + \mathrm{H.c.}\right]\\ 
    & \qquad \qquad \qquad \qquad \qquad+ h\sum_{j=1}^{N}\left(c_j^\dagger c_j - \tfrac{1}{2}\right).
    \label{H_fermion}
\end{aligned}
\end{equation}
}
{\color{black}The charge
\begin{equation}
Q = \sum_{j=1}^{N} c_j^\dagger c_j = \sum_{j=1}^{N}\frac{1-\sigma_j^z}{2}
\label{eq:charge}
\end{equation}
commutes with every string $\mathcal{Z}_{j,r}$. For $\gamma=0$ the Hamiltonian
conserves $Q$ and is invariant under the $U(1)$ rotations $e^{i\phi Q}$. For
$\gamma \neq 0$, the pairing terms break this symmetry.

\subsection{Ground state and spectrum}}
\label{sec:model_spectrum}
{\color{black}Introducing momentum modes
$d_k=N^{-1/2} \sum_{j=1}^{N}e^{-ikj}c_j$ and the Fourier transform of the couplings
\begin{equation}
\tilde{J}^{\alpha}_k = J \sum_{r=1}^{N_c} \frac{e^{ikr}}{r^\alpha}.
\label{eq:Jk}
\end{equation}
We can express the Hamiltonian
in terms of the Nambu spinor $\Psi_k=(d_k,d_{-k}^\dagger)^{T}$ as, 
\begin{equation}
H = \sum_{k>0}\Psi_k^\dagger
    \bigl[\,(h-\mathrm{Re}\,\tilde{J}^{\alpha}_k)\,\tau^z + \gamma\,\mathrm{Im}\,\tilde{J}^{\alpha}_k\,\tau^y\bigr]
    \Psi_k ,
\label{eq:H_Nambu}
\end{equation}
with $\tau^{y,z}$ Pauli matrices in Nambu space.} This is followed by a Bogoliubov transformation,
\begin{equation}
\begin{pmatrix}
\eta_k \\
\eta_{ - k}^\dagger
\end{pmatrix}
=
\begin{pmatrix}
\cos\left(\Delta_k/2\right) & -i \sin\left(\Delta_k/2\right) \\
-i \sin\left(\Delta_k/2\right) & \cos\left(\Delta_k/2\right)
\end{pmatrix}
\begin{pmatrix}
d_k \\
d_{ - k}^\dagger
\end{pmatrix}.
\end{equation}
This allows the Hamiltonian to assume a diagonal form as
\begin{equation}
H = \sum_k \varepsilon_k \left( \eta_k^\dagger \eta_k - \frac{1}{2} \right).
\end{equation}
The quasiparticle dispersion relation reads $\varepsilon_k =
\sqrt{\left(h - \mathrm{Re}\,\tilde J_k^\alpha\right)^2 + \left
(\gamma\,\mathrm{Im}\,\tilde J_k^\alpha\right)^2}$.
The Bogoliubov angle $\Delta_k$ is defined through
\begin{equation}
\cos(\Delta_k) = \frac{h - \mathrm{Re}\,\tilde J_k^\alpha}{\varepsilon_k},
\qquad
\sin(\Delta_k) = \frac{\gamma\,\mathrm{Im}\,\tilde J_k^\alpha}{\varepsilon_k}.
\label{bogo-angle}
\end{equation}
In the isotropic limit $\gamma = 0$, the Hamiltonian reduces to the long-range XX model, which possesses a global $U(1)$ symmetry. For $\gamma \neq 0$, however, this $U(1)$ symmetry is broken. 

{\color{black} 
The ground state of Eq.~\eqref{H_fermion} is the vacuum of the modes $\eta_k$. In terms
of the momentum modes, it takes the BCS form
\begin{equation}
|\Psi(h,\gamma)\rangle = \prod_{k>0}
\left[\cos\frac{\Delta_k}{2} + i\sin\frac{\Delta_k}{2}\,d_k^\dagger d_{-k}^\dagger\right]
|0\rangle ,
\label{GS}
\end{equation}
where $|0\rangle$
is the fermionic vacuum, $d_k|0\rangle=0$. The ground
state is a superposition of Cooper pairs with momenta $(k,-k)$. For $\gamma \neq 0$, each
pair changes the charge $Q$ by two, and the state breaks the $U(1)$ symmetry.\\

\begin{figure*}[t!]
    \centering
    \includegraphics[scale = 0.45]{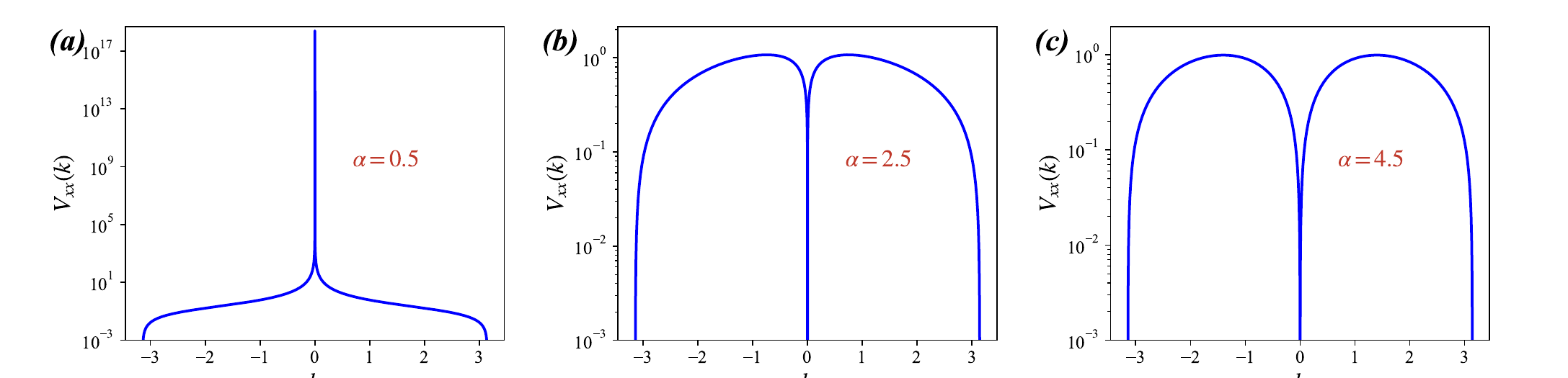}
    \caption{\textit{Quasi-particle velocity in different interaction regime:} The panel $(a),(b)$ and, $(c)$ depicts quasi-particle velocity for $\alpha = 0.5$, $2.5$ and, $4.5$, respectively.}
\label{group-velocity}
\end{figure*}
\subsection{Criticality}

{\color{black}
The critical value of the magnetic field, $h_c$, can be found from the gap closing condition $\epsilon_k=0$. The gap closes only if both $h-\text{Re}(J^{\alpha}_k)$ and  $\gamma\text{Im}(J^{\alpha}_k)$ are zero. For $\gamma\ne0$, $\gamma\text{Im}(J^{\alpha}_k)$ is zero at $k=0,\pm\pi$. At $k=0$, $h-\text{Re}(J^{\alpha}_k)$ can be expanded in the large $N_c$ limit as:
\begin{equation}
    \frac{h_c}{J}=\zeta(\alpha)+\frac{N_{c}^{1-\alpha}}{1-\alpha}+\frac{1}{2}N_c^{-\alpha}+\mathcal{O}(N_c^{-\alpha-1}),
    \label{Euler-Mclaurin}
\end{equation}
whereas at $k=\pm\pi$, the expansion is 
\begin{equation}
  h_c/J=-\eta(\alpha)+O(N_c^{-\alpha}).  
\end{equation}
For $\alpha>1$ and $N_c\to\infty$, we have 
\begin{equation}
    J\sum_{r\le N_c}(1/r^\alpha)=J\zeta(\alpha),
\end{equation}
and
\begin{equation}
    J\sum_{r\le N_c}(-1)^r(1/r^\alpha)=-J\eta(\alpha),
\end{equation}
where $\zeta$ and $\eta$ are the Riemann zeta and Dirichlet eta functions. As at $k=\pm \pi$, we get $h_c<0$, which lies outside the range of field considered in this work. For large values of $\alpha$, ${h_c}/{J}\to1$ as $\alpha \to \infty$. At finite values of $\alpha$, till $\alpha>1$, the critical point grows with decreasing $\alpha$, e.g., ${h_c}/{J} \approx 1.202$ at $\alpha=3$ and  ${h_c}/{J}\approx 2.612$ at
$\alpha=1.5$.
For $\alpha<1$, $h_c/J=\sum_{r\le N_c}r^{-\alpha}$ behaves as
$N_c^{1-\alpha}$ and depends on the cutoff. However, Eq.~(\ref{Euler-Mclaurin}) diverges at $N_c = \infty$, implying the absence of a critical point in the thermodynamic case.

\subsection{Post-quench Hamiltonian and group velocity}
At $\gamma=0$, the Hamiltonian is diagonal in the momentum modes,
\begin{equation}
H_{XX} = \sum_k \epsilon_{XX}(k)\left(d_k^\dagger d_k-\frac{1}{2}\right),
\label{eq:H_XX}
\end{equation}
where $\epsilon_{XX}(k) = h-\mathrm{Re}\tilde{J}^{\alpha}_k$. Throughout this work, we quench to $h=0$, where
$\epsilon_{XX}(k)=-\sum_{r\le N_c}J_r\cos(kr)$. The corresponding group velocity,
\begin{equation}
V_{XX}(k) = \frac{d\epsilon_{XX}(k)}{dk},
\label{VXX}
\end{equation}
is odd. Hence, the two members of a pair $(k,-k)$ propagate in opposite directions with
speeds $|V_{XX}(k)|$. For the remainder of this work, we set $J=1$ for convenience.

{\color{blue}  }
\begingroup

\newcolumntype{M}{>{$\vcenter\bgroup\vskip6pt\hbox\bgroup$\displaystyle}l<{$\egroup\vskip6pt\egroup$}}

{\color{black}
\subsection{Dispersions near $k=0$ and $\pm \pi$}
For $h=\gamma=0$, the dispersion reduces to
\begin{equation}
 \epsilon(k,\alpha,N_c)
=-\text{Re}\tilde J^{\alpha}_k(N_c).
\end{equation}
In the limit $N_c \to \infty$, the coupling can be expressed in terms of the polylogarithm function, $\tilde J^{\alpha}_k \equiv \tilde J^{\alpha}_k(\infty)=\operatorname{Li}_{\alpha}(e^{ik})$.

Using its small-$k$ expansion around $k=0$, we obtain \cite{NIST}
\begin{equation}
\begin{split}
    \epsilon_{k \to 0}(\alpha,\infty) &= -\zeta(\alpha)-C_\alpha|k|^{\alpha-1}\\
    &+\frac{\zeta(\alpha-2)}{2}k^2 -\frac{\zeta(\alpha-4)}{24}k^4+\cdots,
\end{split}
    \label{smallk-dispersion}
\end{equation}
where
\begin{equation}
    C_\alpha=\Gamma(1-\alpha)\sin\left(\frac{\pi\alpha}{2}\right).
    \label{dispersion-coeff}
\end{equation}
This expression applies for generic $\alpha>0$, excluding positive odd integers, where the corresponding limits must be taken. In particular, at $\alpha = 1$ and $3$, the divergent coefficients combine to produce a logarithmic correction. The leading order behaviour turns out to be 
\begin{equation}
\epsilon_{k \to 0}(\alpha=3,\infty)
=-\zeta(3)-\frac{k^2}{2}\ln|k|
+\frac{3k^2}{4}+\frac{k^4}{288}
+O(k^6),
\end{equation}
and
\begin{equation}
\epsilon_{k \to 0}(\alpha=1,\infty)
=+\ln|k|-\frac{k^2}{24}-\frac{k^4}{2880}
+O(k^6).
\end{equation}

Hence, the dispersion is nonanalytic in the leading order for every $\alpha<3$: for $1<\alpha<3$, it is proportional to $|k|^{\alpha-1}$, has a logarithmic dependence $|k|^2\ln|k|$, at $\alpha =3$, and is quadratic for $\alpha>3$.
For $\alpha\le1$, the thermodynamic-limit coupling diverges at $k=0$, as
$|k|^{\alpha-1}$ (as $\ln|k|$ at $\alpha=1$). This divergence is removed by a finite
interaction cutoff. For $|k|\ll1/N_c$ the finite sum can be expanded as
\begin{equation}
\epsilon_{k\to 0}(\alpha,N_c)
= -H_{N_c}^{(\alpha)}
+\frac{k^2}{2}H_{N_c}^{(\alpha-2)}-\frac{k^4}{24}H_{N_c}^{(\alpha-4)}
+O(k^6),
\label{cutoff_dis}
\end{equation}
where the generalized harmonic number, $H_{N_c}^{(\alpha)}$,can be expanded in large $N_c$ as,
\begin{equation}
H_{N_c}^{(\alpha)}
=\zeta(\alpha)+\frac{N_c^{1-\alpha}}{1-\alpha}
+\frac{1}{2N_c^{\alpha}}
-\frac{\alpha}{12N_c^{\alpha+1}}+\cdots. 
\end{equation}
Consequently, for $0<\alpha<1$, we obtain the leading order term of the dispersion
\begin{equation}
\epsilon_{k\to0}(\alpha,N_c)=-\frac{N_c^{1-\alpha}}{1-\alpha}-\zeta(\alpha)-\frac{1}{2N_c^{\alpha}}+\mathcal{O}(N_c^{-\alpha-1}),
\end{equation}

At $k=\pm\pi$, $\tilde J_{\pm\pi}(\alpha,\infty)
=-\eta(\alpha)$. Expanding in this small neighbourhood, we obtain
\begin{equation}
\begin{split}
\epsilon_{k\to \pm\pi}(\alpha,\infty)
&=\eta(\alpha)
-\frac{\eta(\alpha-2)}{2}(k\mp\pi)^2\\&+\frac{\eta(\alpha-4)}{24}(k\mp \pi)^4
+O((k\mp\pi)^6).
\end{split}
\label{pi-dispersion}
\end{equation}
This expansion remains analytic for all $\alpha>0$, including $\alpha<1$. Hence, in contrast to the behaviour near $k=0$, no finite interaction cutoff is required to obtain a well-defined thermodynamic-limit dispersion around $k=\pm\pi$.

{\color{black} \subsection{Quasi-particle Group Velocity near $k=0$ and $k=\pm \pi$}
\label{Velocity_asymptotics}
In the limit , $N_c \rightarrow \infty$, the group velocity $V_{XX}(k)$ too, can be written in terms of polylogarithm function, $V_{XX}(k)=\text{ Im}(\text{Li}_{\alpha-1}(e^{ik}))$.

\begin{figure}[t]
    \centering
    \includegraphics[width=\columnwidth]{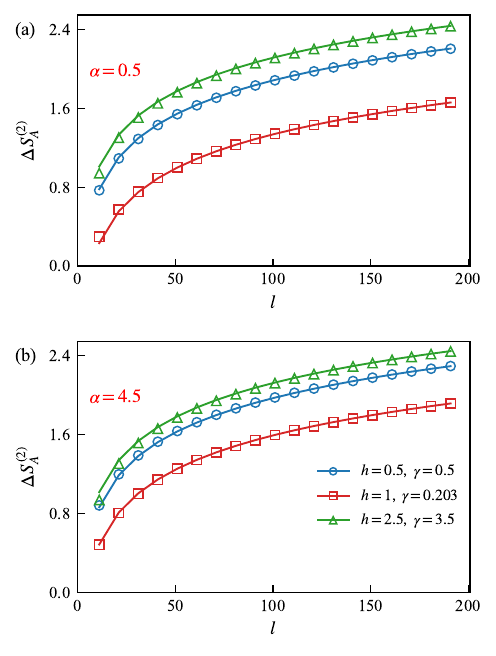}
    \caption{ Scaling of $\Delta S^{(2)}_{A}$  with the subsystem size $l$ for (a) long-range  ($\alpha=0.5$) regime and (b) short-range ($\alpha=4.5$) interactions. Symbols correspond to the direct numerical result, while the solid lines correspond to the result obtained via Eq.~(\ref{Asymmetry_Scaling}).}
    \label{fig: GS_scalling}
\end{figure}
Depending on $\alpha$ and its cut-off range, the quasi-particle velocity at $k=0$ can be finite or can diverge to infinity. The small-$k$ behavior is directly followed from the Eqs.~(\ref{smallk-dispersion}-\ref{dispersion-coeff}) as, 
\begin{equation}
    \begin{aligned}
    V_{XX}(\alpha, \infty)\big|_{k \to 0}
    =&-(\alpha-1)C_{\alpha}
    \text{sgn}(k)|k|^{\alpha-2}\\
    &+\zeta(\alpha-2)k
    -\frac{\zeta(\alpha-4)}6 k^3+\cdots.
    \end{aligned}
\end{equation}
 This expression is valid for every $\alpha>0$ except the odd integers, $\alpha\geq3$. For $\alpha>3$, $ V_{XX} \sim \zeta(\alpha-2)k$ and thus has a linear dependence on $k$ in the leading order and at $\alpha=3$, it turns out to be $V_{XX} \sim -k \ln |k|+k$. The leading order behavior for $\alpha<3$ deviates from linearity, $V_{XX} \sim -(\alpha-1)C_\alpha \text{sgn}(k)|k|^{\alpha-2}$. For $\alpha<2$, $V_{XX}$ diverges at $k=0$  and $N_c\to\infty$. This is regularised via an interaction cut-off by keeping $N_c$ finite.

For $\alpha=2$, we obtain 
\begin{equation}
    V_{XX}(\alpha=2, \infty)=\frac{1}{2}\text{sgn}(k)(\pi- |k|).
\end{equation}
{\color{black}
However, for $\alpha>2,\alpha \in \mathbb{Z}^{+}$ the expansion around $k=0$ is,
\begin{equation}
\begin{split}
V_{XX}(\alpha,\infty)\big|_{k\to 0}=&\sin\!\Big(\frac{\pi(\alpha-2)}{2}\Big)\frac{k^{\alpha-2}}{(\alpha-2)!}
\big(H_{\alpha-2}-\ln|k|\big)\\
&+\cos\!\Big(\frac{\pi(\alpha-2)}{2}\Big)\frac{\pi\,\mathrm{sgn}(k)\,|k|^{\alpha-2}}{2\,(\alpha-2)!}\\
&+\sum_{\substack{j\ \mathrm{odd}\\ j\neq\alpha-2}}(-1)^{\frac{j-1}{2}}\,
\zeta(\alpha-1-j)\,\frac{k^{j}}{j!},
\end{split}
\end{equation}

Here, $H_{n}=\sum_{m=1}^{n}1/m$ is the harmonic number ($H_0=0$). This expression can be used to find the leading-order behaviour of $V_{XX}$ for odd integer values of $\alpha$. At $k=\pm \pi$, $V_{XX}=0$. Around $k=\pm \pi$ we can write the group velocity in a small neighbourhood around the left and right of $\pm\pi$, respectively, 
\begin{equation}
\begin{split}
V_{XX}&(\alpha,\infty)\big|_{k\to\pm\pi}=-(k\mp\pi)\,\eta(\alpha-2)\\
&+\frac{(k\mp\pi)^3}{6}\,\eta(\alpha-4)-\frac{(k\mp\pi)^5}{120}\eta(\alpha-6)\\
&+\cdots,
\end{split}
\label{vxx-pi}
\end{equation}
which is valid for every value of $\alpha$. This expression remains well-defined and finite for any $\alpha$, around $k=\pm \pi$.} 

Fig.~\ref{group-velocity} shows the quasiparticle velocity $V_{XX}(k)$ as a function of $k$ for three representative values of the power-law decay exponent $\alpha$. For $\alpha=0.5$, the velocity diverges near $k=0$, while it vanishes at $\pm \pi$, indicating the presence of fast modes with arbitrarily large velocities at small momenta and slow modes near $k=\pm\pi$.  In contrast, for $\alpha=2.5$ and $4.5$, the velocity remains finite everywhere and vanishes at $k=0$ and $k=\pm\pi$. Thus, $\alpha$ controls the low-momentum behaviour of the quasi-particles, and eventually plays a crucial role in deciding the fate of the QME.

The group velocity comes into play later in the dynamics through the quasiparticle description presented in Sec.~\ref{quasi-particle}, and only the slow quasiparticles with near-zero values control the late-time asymmetry that finally decides the QME. For $\alpha<2$, the thermodynamic-limit velocity diverges as $k \to 0$. The modes near $k=0$ are therefore the fastest ones, as they leave the subsystem early and do not contribute to the late-time asymmetry. The
slow modes reside near $k=\pm\pi$, and are finite for every $\alpha>0$, whereas for $\alpha>2$ the modes near $k=0$ are slow as well. The late-time expressions of Sec.~\ref{quasi-particle} are accordingly derived from the thermodynamic forms of the velocity near only those points that host slow modes, whereas the exact charged-moment results for the subsystems with size $l$ are obtained with a finite, but large, cut-off $N_c$. As we shall present later, the two agree at late times for all $\alpha$ considered.

\begin{figure*}[t!]
    \centering
    \includegraphics[scale = 0.48]{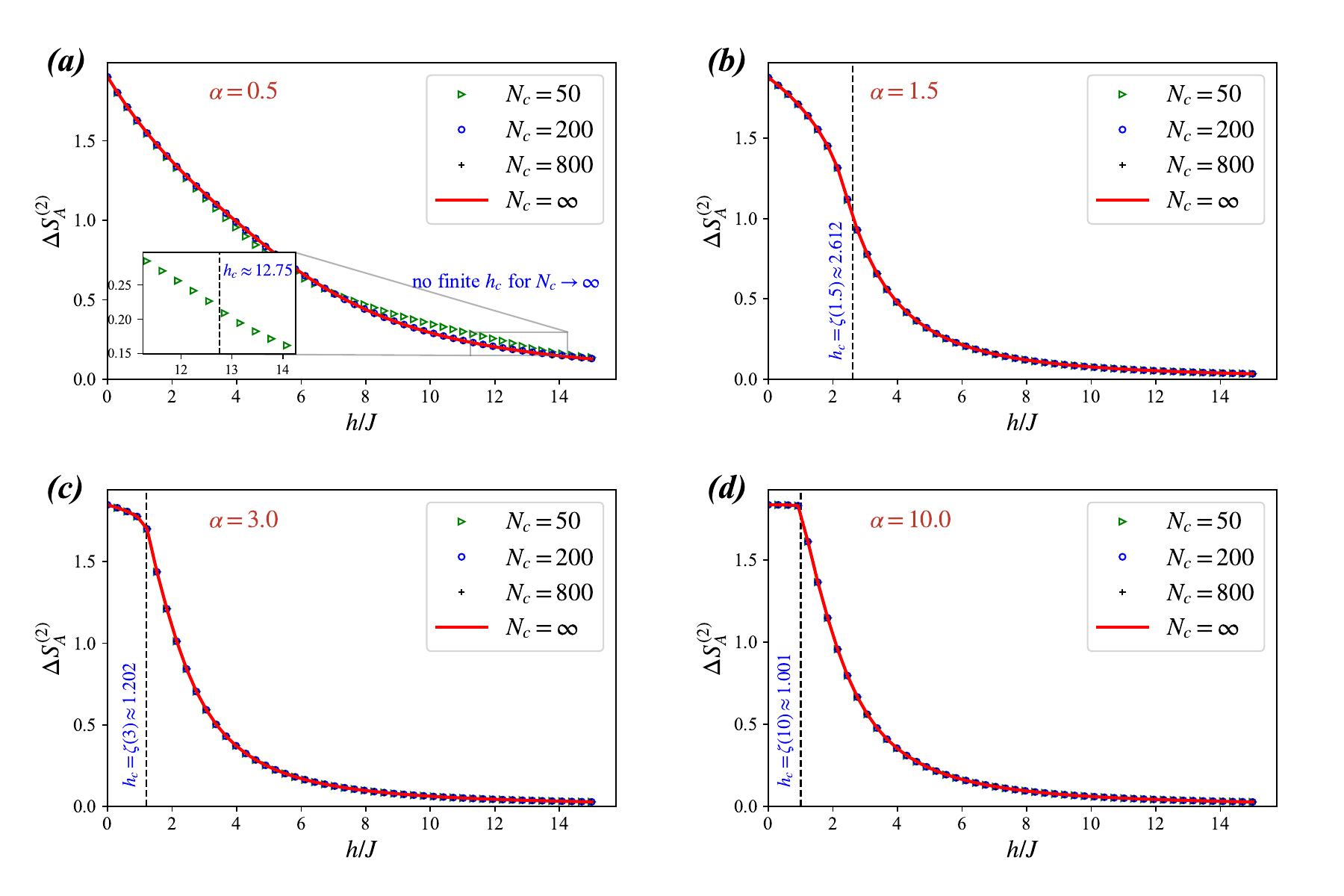}
    \caption{ $\Delta S_A^{(2)}$ versus $h$ is depicted for different long-range interactions for the ground state with $\gamma = 1$, and the subsystem size is $l=50$. (a) and (b) for $\alpha = 0.5$ and $\alpha = 1.5$, respectively; the asymmetry starts decaying monotonically with $h$. In (c) and (d) for the short-range ($\alpha \geq 3$), the asymmetry is independent of $h$ till $h/J\approx1$ and monotonically decays to zero with increasing $h$. For all the cases in the large cutoff, the $\Delta S_A^{(2)}$ converge. For $\alpha=1.5,3.0,10.0$ the critical value of magnetic field can be calculated using $\frac{h_c}{J}=\zeta(\alpha)$ in $N_c \to \infty$ limit. For $\alpha=0.5$, there is no critical point in the limit $N_c \xrightarrow{} \infty$. However, there is a cutoff-dependent critical point at finite $N_c$ as given by Eq. (\ref{Euler-Mclaurin}). We are considering $h>0$ in this work.}
    \label{fig:static_EE}
\end{figure*}

\section{Entanglement asymmetry}
\label{sec_EA}
In order to quantify the extent of symmetry breaking and to investigate the associated symmetry restoration dynamics in our systems, we make use of the entanglement asymmetry (EA) ~\cite{ares2023entanglement,Murciano2024,Ares2025,Lastres_2025,PhysRevLett.133.010401}. EA for a subsystem $A$ with conserved charge $Q$ is defined as,
\begin{equation}
\Delta S_A = S(\rho_{A,Q}) - S(\rho_A),
\quad 
S(\rho) = -\operatorname{Tr}(\rho \log \rho),
\end{equation}
where $\rho_A = \mathrm{Tr}_{\bar A}(|\psi\rangle\langle\psi|)$ is the reduced density matrix of the subsystem $A$ obtained by tracing out its complement $\bar A$. The operator $\rho_{A,Q}$ denotes the projected density matrix, obtained by decomposing $\rho_A$ into the charge sectors of the operator $Q_A$, $\rho_{A,Q} = \sum_{q \in \mathbb{Z}} \Pi_q \, \rho_A \, \Pi_q$,
where $\Pi_q$ is the projector onto the eigenspace of $Q_A$ with eigenvalue $q$. A vanishing EA implies that $[\rho_A, Q_A] = 0$, i.e., the subsystem $A$ preserves the symmetry. {\color{black}A larger value of $\Delta S_A$ indicates stronger symmetry breaking, whereas a smaller value signals closer adherence to the symmetry. The R{\'e}nyi entanglement
asymmetry is
\begin{equation}
\Delta S_A^{(n)} = S^{(n)}(\rho_{A,Q}) - S^{(n)}(\rho_A),
\label{REA_def}
\end{equation}
where $S^{(n)}(\rho)=(1-n)^{-1}\ln\mathrm{Tr}\rho^{n}$ is the R{\'e}nyi entropy and it reduces to the von Neumann entropy for $n\to1$.}

To compute $\Delta S_A$, we introduce another quantity called charged moments \cite{Murciano2024}
\begin{equation}
Z_{n}(\boldsymbol{\tilde{\alpha}}) =
\mathrm{Tr}\!\left(
\prod_{j=1}^{n}
\rho_A \,
e^{i(\tilde{\alpha}_j-\tilde{\alpha}_{j+1})Q_A}
\right),
\qquad
\boldsymbol{\tilde{\alpha}} = (\tilde{\alpha}_1,\ldots,\tilde{\alpha}_n),
\label{eq:charged_moments1}
\end{equation}
with the periodic identification $\tilde{\alpha}_{n+1} \equiv \tilde{\alpha}_1$. In terms of these charged moments, EA can be expressed as
\begin{equation}
\Delta S^{(n)}_A =
\frac{1}{1-n}
\log\!\left[
\int_{-\pi}^{\pi}
\frac{d\tilde{\alpha}_1 \cdots d\tilde{\alpha}_n}{(2\pi)^n}
\frac{Z_n(\boldsymbol{\tilde{\alpha}})}{Z_n(\boldsymbol{0})}
\right].
\label{eq:entanglement_asymmetry_charged}
\end{equation}
If $[\rho_A, Q_A]=0$, then $Z_n(\boldsymbol{\tilde{\alpha}})=Z_n(\boldsymbol{0})$, implying $\mathrm{Tr}\,\rho^n_{A,Q}=\mathrm{Tr}\,\rho^n_A$ and hence $\Delta S^{(n)}_A=0$.

{\color{black}All the states considered here are Gaussian states. Hence,  $\rho_A$ and the charged moments can be determined from the two-point correlations.} 
With a change of variable, $\tilde \alpha=\tilde{\alpha_1}-\tilde{\alpha_2}$ in Eq.~(\ref{eq:charged_moments1}), the charged moment for $n=2$ can be expressed in terms of the modified two-point correlation matrix $\Gamma_{\tilde \alpha}=\Gamma e^{i\tilde\alpha n_{A}}$ as \cite{Murciano2024},
\begin{equation}
Z_{2}(\tilde{\alpha}) =
\det\!\left(
\frac{\mathbb{I} + \Gamma_{\tilde{\alpha}} \Gamma_{-\tilde{\alpha}}}{2}
\right)^{1/2}.
\label{eq:Zn_correlation}
\end{equation}
For a subsystem of length $l$, the matrix elements of $\Gamma_{\tilde\alpha}$ are given by

\begin{equation}
(\Gamma_{\tilde\alpha})_{j j'} =
\int_{-\pi}^{\pi} \frac{dk}{2\pi}
G_{\tilde\alpha}(k) e^{-ik(j-j')},
\qquad
j,j' = 1,\ldots,l .
\label{topl}
\end{equation}
with
\begin{equation}
G_{\tilde\alpha}(k) =
\begin{pmatrix}
e^{i\tilde{\alpha}}\cos\Delta_k & -i e^{-i\tilde{\alpha}}\sin\Delta_k \\
i e^{i\tilde{\alpha}}\sin\Delta_k & -e^{-i\tilde{\alpha}}\cos\Delta_k
\end{pmatrix},
\label{Green_alpha}
\end{equation}
{\color{black}where $\Delta_k$ is the Bogoliubov angle of the initial state.}
For a finite $N$, the integration in Eq.~(\ref{topl}) is replaced by a discrete summation,
\begin{equation}  (\Gamma_{\tilde\alpha})_{jj'}=\frac{1}{N}\sum_{m=-N/2}^{N/2-1}G_{\tilde\alpha}(k_m)e^{-ik_m(j-j')},
\end{equation}
where $ k_m =\frac{2\pi}{N}\Big(m+\frac{1}{2}\Big)$ and $j,j'=1,\dots,l.$. For $n=2$, substituting this expression into Eq.~(\ref{eq:entanglement_asymmetry_charged}), one obtains a tractable formulation of EA {\color{black}that requires calculating determinant of a $2l\times2l$ matrix and a one-dimensional integral
over $\tilde\alpha$.
\begin{equation}
 \Delta S^{(2)}_A =-\log\Big(\int^{\pi}_{-\pi} \frac{d\tilde{\alpha}}{2\pi}\frac{Z_2(\tilde{\alpha})}{Z_2(0)}\Big)
 \label{numerical_asymmetry}
\end{equation}
} \\

\subsection{{\color{black}Ground-state asymmetry and criticality}}
{\color{black} Before turning to the dynamics, we characterize the asymmetry of the ground
states Eq.~(\ref{GS}). This allows one to establish a hierarchy among the symmetry-broken states, thereby identifying a pair that satisfies the first condition of the QME.}
We present $\Delta S_A^{(2)}$, as a function of $h$ in Fig.~\ref{fig:static_EE} for the four exponents $\alpha=0.5$, $1.5$, $3$, and $10$.

{\color{black} For any given $\alpha$, the ground state EA, $\Delta S_{A}^{(2)}$ of this model scales with subsystem size $l$ according to the scaling relation \cite{Murciano2024}, (see in Fig.~ \ref{fig: GS_scalling}).
\begin{equation}
    \Delta S^{(2)}_A 
    = \frac{1}{2}\log (\frac{\pi\, g(\alpha,\gamma,h)l}{2})
    + \mathcal{O}(l^{-1}),
    \label{Asymmetry_Scaling}
\end{equation}
where 
\begin{equation}
    g(\alpha,\gamma,h) = \int_{-\pi}^{\pi}\frac{dk}{2\pi}\sin^{2}(\Delta_k). 
    \label{eq_g}
\end{equation}
$g$ is the pairing amplitude averaged over the Brillouin zone and can be physically interpreted as the charge fluctuation of the subsystem. 
The validity of Eq.~(\ref{Asymmetry_Scaling}) relies on the Gaussian saddle-point approximation, as discussed in Appendix~\ref{Scaling of EA}. Therefore, Eq.~(\ref{Asymmetry_Scaling}) holds for $lg\gg1$. Outside of this regime, $\Delta S_A^{(2)}$ can be obtained by evaluating the $\tilde\alpha$ integral in Eq.~(\ref{analytic_EA_2}) numerically (see Appendix \ref{Scaling of EA}). 

The interaction range parameter, $\alpha$, modifies the Bogoliubov angle, $\Delta_k$. In Fig.~(\ref{fig: GS_scalling}), the scaling relation of EA in Eq.~(\ref{Asymmetry_Scaling}) is plotted against the exact numerical value obtained using Eq.~(\ref{numerical_asymmetry}) for three different $(h,\gamma)$ pairs, $(h,\gamma)=(0.5,0.5),(1.0,0.203),(2.5,3.5)$. Fig.~(\ref{fig: GS_scalling})(a), shows how EA scales with subsystem size $l$ for $\alpha=0.5$ and Fig.~(\ref{fig: GS_scalling}(b)) shows the same for $\alpha=4.5$. In all cases, the entanglement asymmetry exhibits a logarithmic growth with subsystem size, with parameter-dependent coefficients. We find excellent agreement between the numerical results and the analytical scaling relation provided in Eq.~(\ref{Asymmetry_Scaling}), and thus confirm its validity across all interaction regimes. 

\par The critical point, $h_c/J=\sum_{r\le N_c}r^{-\alpha}$, at which the gap closes, distinguishes the behavior of $g$ in either phase. In the limit $\alpha \to \infty$, the model is reduced to the nearest-neighbor XY chain, for which $h_c/J=1$, and $g=\gamma/(1+\gamma)$ if $h < h_c=1$, and hence is essentially independent of the field, whereas it assumes the value $g=\left({\gamma^2}/{(1 - \gamma^2)}\right) \left( {|h|}/{(\sqrt{h^2 + \gamma^2 - 1)}} - 1 \right)$ for $h > h_c=1$. The asymmetry is therefore nearly constant up to $h=h_c$ at high $\alpha$ values, beyond which it displays rapid decay, as demonstrated in the Fig.~\ref{fig:static_EE}(d) for $\alpha=10$, where  
$h_c/J= \zeta(10) \simeq 1.001$.

At finite and intermediate values of $\alpha$, the plateau acquires a slope and the kink shifts to higher values with decreasing $\alpha$, with criticality set as $h_c=J\zeta(\alpha)$. As shown in the Figs.~(b-c), for $\alpha=3$ the asymmetry decreases slowly up to
$h_c/J \approx 1.202$, and for $\alpha=1.5$ up to
$h_c/J \approx 2.612$, beyond which fast decay is again observed in both cases.

In contrast, for the genuinely long-range case ($\alpha=0.5$), as shown in Fig.~\ref{fig:static_EE}(a),  $\Delta S_A^{(2)}$ decreases monotonically starting from $h=0$ and eventually approaches zero as $h$ becomes large in the limit $N_c \to \infty$, where the system is devoid of a phase transition. There is, however, a cutoff-dependent criticality, e.g., $h_c/J \approx 12.75$ for $N_c=50$. Convergence to the infinite-cutoff limit is examined in the Figs.~\ref{fig:static_EE}(a-d). For $\alpha>1$, the results for $N_c=50$, $200$, and $800$ coincide with the limit $N_c\to\infty$ on the scale of the figure. In the long-range case, small cut-off dependence can be noticed within the range of $N_c$ considered in this work. 
}

\section{Quantum Mpemba Effect} 
\label{QME}
{\color{black}Consider two initial states, 1 and 2, evolved with the same Hamiltonian. The QME occurs if
\begin{equation}
\begin{aligned}
&\text{(i)}\quad \Delta S_{A,1}^{(2)}(0) > \Delta S_{A,2}^{(2)}(0),\\
&\text{(ii)}\quad \Delta S_{A,1}^{(2)}(t) < \Delta S_{A,2}^{(2)}(t)
   \quad\text{for all } t>t_M ,
\end{aligned}
\label{QME_def}
\end{equation}
where the Mpemba time $t_M$ is the time of the last crossing of the two curves. The asymmetries of two states can cross more than once~\cite{Chalas_2024}. As a result, a single crossing at a finite time does not by itself establish the effect. We encounter such transient crossings as well. Condition (ii) fixes such a multi-crossing situation. These relations imply that the state that is initially further from equilibrium reaches the stationary regime faster, which characterizes the QME.}

In this section, we present a detailed analysis of the EA for two classes of initial states, the tilted ferromagnetic state and the ground-state, and examine the occurrence of QME across different interaction regimes.

\begin{figure*}[t!]
    \centering
    \includegraphics[scale = 0.44]{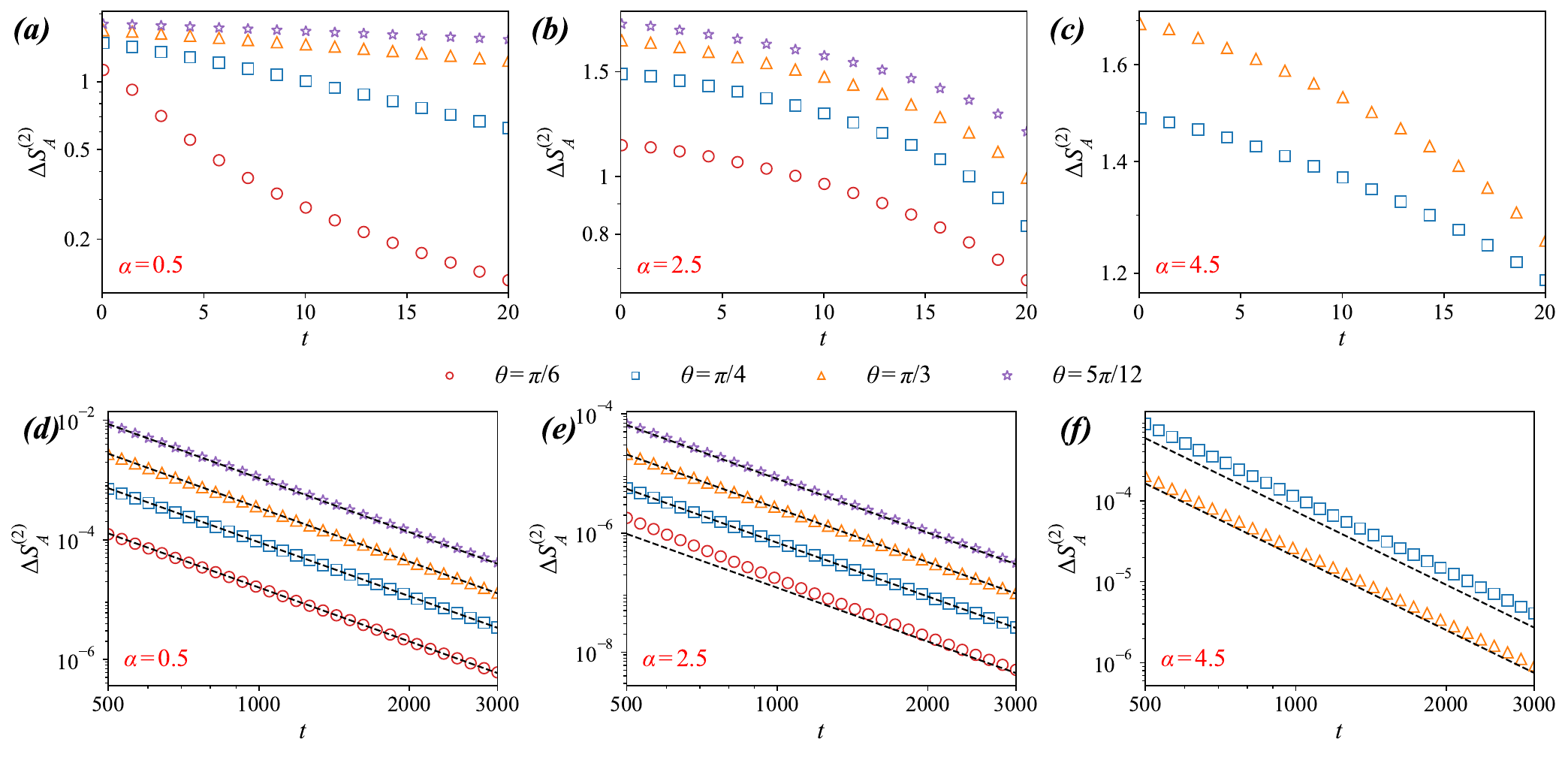}
    \caption{$\Delta S_A^{(2)}$ after a quench for tilted ferromagnetic
states with different tilt angles $\theta$, for $\alpha=0.5$, $2.5$ and $4.5$
(left to right). The top row, (a)--(c), shows early-time behavior while the bottom row, (d)--(f), shows late-time behavior of EA. The dashed line in the bottom row is the EA as calculated using the quasi-particle picture in the large-time limit. For $\alpha=0.5$ and $2.5$ the initial ordering of EA does not get altered at late times, so no QME occurs [(a), (b), (d), (e)]. This is in agreement with the condition given in Sec.~\ref{Tilted state QPP} that for a pair $(\theta_1,\theta_2)$ with $\theta_1+\theta_2<\pi$, the QME is absent in the  regime $\alpha<\alpha_c=3$. For $\alpha=4.5$, the ordering is reversed at late times [(c), (f)], showing QME for the given pair of states. For $\alpha=4.5$, the pair $\theta_1=\pi/4$ and $\theta_2=\pi/3$ displays the QME, since both angles lie below the threshold tilt, $\theta_1<\theta_2<\theta^*(4.5)\approx0.448\pi$. Here, the subsystem size is taken to be $l=50$.}
    \label{fig:QME_tilted_st}
\end{figure*}

{\color{black}\subsection{Initial states and driving Hamiltonian}

We consider two classes of the initial states, $\psi_{\text in}$, without $U(1)$ symmetry, 
i.e., $[\rho,Q]\neq0$, where $\rho=|\psi_{\text in}\rangle\langle\psi_{\text in}|$. At time $t=0$, the system is quenched to $H_{XX}$ at $h=0$. In the following analysis, we investigate two distinct classes of the initial states: 

(i) The first class of the initial states is tilted ferromagnetic states, defined as
\begin{equation}
|\Psi(\theta)\rangle =
\frac{|\theta;\nearrow\nearrow\cdots\rangle-|-\theta;\nearrow\nearrow\cdots\rangle}{\sqrt{2}},
\label{tilted}
\end{equation}
where $|\theta;\nearrow\nearrow\cdots\rangle = e^{i\frac{\theta}{2}\sum_j\sigma_j^y}
|{\uparrow\uparrow\cdots}\rangle$.  For $\theta \neq m\pi$ ($m \in \mathbb{Z}$), the tilted state breaks the $U(1)$ symmetry associated with the conserved charge $Q$.  This is a Gaussian state of the form given in Eq.~(\ref{GS}) with
\begin{equation}
\cos\Delta_k = \frac{2\cos\theta-(1+\cos^2\theta)\cos k}{1+\cos^2\theta-2\cos\theta\cos k},
\end{equation}
Note that the tilted states are independent of $\alpha$. The interaction range enters its dynamics only through $\epsilon_{XX}(k)$.  EA of a tilted ferromagnetic state increases monotonically with the tilt angle $\theta$ up to $\theta=\pi/2$, then it decreases with $\theta$ and eventually goes to zero as $\theta \to \pi$. In the initial states, EA is given by, 
\begin{equation}
    \Delta S^{(2)}_A 
    = \frac{1}{2}\log \left(\frac{\pi\, \sin^2(\theta)l}{4}\right)
    + \mathcal{O}(l^{-1}).
    \label{Asymmetry_Scaling_tilt}
\end{equation}

(ii) The second class of initial states that we investigate are the ground states  $|\Psi(h,\gamma)\rangle$ (see Eq.~\eqref{GS}) of the Hamiltonian in Eq.~\eqref{Hamiltonian} with $\gamma \ne 0$. Unlike the tilted states, this class of initial states is dependent upon the interaction range controlled by $\alpha$.

The quenched system restores $U(1)$ symmetry at long times. The driven system settles to a generalised Gibbs ensemble of conserved occupations that commute with $Q$.  
}

\subsection{Tilted states}
{\color{black}We first consider the initial state as the tilted ferromagnetic state in Eq.~\eqref{tilted}.}  For this initial state, $|\Psi (\theta)\rangle$, we study the time evolution of $\Delta S_A^{(2)}$ through quench dynamics, with the post-quench hamiltonian corresponds to $h=0,\gamma=0$. {\color{black}Fig.~\ref{fig:QME_tilted_st} shows the time evolution of the EA for different tilt angles and interaction strength $\alpha$, with subsystem size $l=50$. {\color{black} In the long-range regime ($\alpha<3$), the initial order of EA remains unchanged in the asymptotic limit. Hence, in this regime, the QME does not occur for the chosen set of tilted initial states. The representative cases for $\alpha=0.5,2.5$ are shown in Figs.~\ref{fig:QME_tilted_st}(a),(d) and \ref{fig:QME_tilted_st}(b),(e) respectively. For $\alpha=0.5$, Fig.~\ref{fig:QME_tilted_st}(a) shows the early-time behaviour of EA for different tilted initial states, and Fig.~\ref{fig:QME_tilted_st}(d) shows the late-time behavior of EA. It is evident from the plots that the initial order EA remains unchanged at large times. Fig. \ref{fig:QME_tilted_st}(b) and  Fig.~\ref{fig:QME_tilted_st}(e) shows early-time and late-time behavior of EA for $\alpha=2.5$. For this case too, the initial ordering of EA does not change at asymptotically large time.}}

In contrast, in the regime of short-range interaction ($\alpha>3$),  for the chosen pair of initial states, the initial order of EA is reversed at late time, signalling the occurrence of the QME. {\color{black}The representative case for $\alpha=4.5$ is displayed in the Figs.~\ref{fig:QME_tilted_st}(c), (f). Fig.~\ref{fig:QME_tilted_st}(c) shows the initial-time dynamics of EA, and Fig.~\ref{fig:QME_tilted_st}(f) shows the late-time behavior of EA.} The state with larger initial EA, corresponding to $\theta=\pi/3$ (yellow triangle), relaxes faster than the state with $\theta=\pi/4$ (blue squares), which has a smaller value of EA initially. The initial order of EA is reversed at asymptotically large time, thus giving the signature of the QME.  In the $\alpha>3$ regime, there can be pairs of initial states for which the initial order of EA remains unchanged at late times. The condition for the occurrence or absence of the QME is followed below within the quasiparticle picture.

For the cases of the tilt angles considered here, although the QME behavior is observed in the short-range interaction regime, the long-range interactions suppress the phenomenon, and consequently, symmetry restoration occurs without reversal of ordering. Importantly, the tilted states are independent of $\alpha$. Hence, the distinctive behavior, i.e., the manifestation of the QME phenomena for the representative states in the short-range systems and its absence in the long-range regime, entirely stems from the post-quench dynamics. Importantly, the considered angles satisfy a criterion,  $\theta_1+\theta_2<\pi$. The origin of this criteria is discussed in Sec.~\ref{quasi-particle}. Pairs that do not satisfy this criteria, i.e., pairs for which
$\theta_1+\theta_2>\pi$,  display the QME in all interaction ranges, and are also discussed in Sec.~\ref{quasi-particle}.

\subsection{Ground states} 
{\color{black} Next, we consider the ground states of the Hamiltonian with $\gamma \ne 0$ (see Eq.~(\ref{Hamiltonian})) as the initial states. Unlike the tilted states, the correlations in the ground states are directly controlled by the interaction-range parameter, $\alpha$, as suggested in Eq.~\eqref{bogo-angle}. The $\alpha$ influences both the initial state structure and the post-quench Hamiltonian.}

In Fig.~\ref{fig:Ent_asymmetry}, we present the time evolution of $\Delta S_A^{(2)}$ following a sudden quench, where the post-quench Hamiltonian again corresponds to the parameters $\gamma = 0$ and $h = 0$. We find that the QME occurs in all ranges of $\alpha$, both short- and long-range regimes, for suitable choices of a pair of initial states.  We first consider the long-range interaction regime defined by $0 < \alpha < 1$, and take $\alpha = 0.5$ as a representative example. Fig.~\ref{fig:Ent_asymmetry}(a) shows the early time dynamics of EA whereas  Fig.~\ref{fig:Ent_asymmetry}(d) depicts the late time behavior of $\Delta S_A^{(2)}$ for three different initial states fixed by the parameters $h \text{ and }\gamma$: $\gamma = 0.203$, $h = 1.0$ (blue circle), $\gamma = 0.585$, $h = 3.11$ (red square), and $\gamma = 0.7$, $h = 0.3$ (green triangle). The results are obtained with the subsystem size is fixed at $l = 50$. For the pairs, $\gamma = 0.203$, $h = 1.0$ (blue circle) and $\gamma = 0.585$, $h = 3.11$ (red square), the corresponding curves intersect at a finite time. The initial order of EA is reversed at finite time, and this reversed order is maintained at late times, as evident from Fig.~\ref{fig:Ent_asymmetry}(d). However, this behavior is not universal for all choices of initial parameters. For example, $\Delta S_A^{(2)}$ curves for the pair $\gamma = 0.7$, $h = 0.3$ (green triangle) and $\gamma = 0.585$, $h = 3.11$ (red square) do not show order reversal of EA at late times. Therefore, no QME is observed for this pair of initial conditions.

Next, we analyze the quasi-long-range regime by considering $\alpha = 2.5$ in Figs.~\ref{fig:Ent_asymmetry}(b) and ~\ref{fig:Ent_asymmetry}(e). In this case, we again find QME for certain initial states. Specifically, the pair $\gamma = 0.203$, $h = 1.0$  and $\gamma = 0.585$, $h = 3.11$ shows  reversal of EA ordering in late time. On the other hand, for a pair of initial states, namely $\gamma = 0.585$, $h = 3.11$ and $\gamma = 0.7$, $h = 0.3$, the initial ordering of EA remains unaltered at late times. Finally, we perform a similar analysis in the short-range regime by fixing $\alpha = 4.5$ (see Fig.~\ref{fig:Ent_asymmetry}(c),(f)). For this case, the initial state pair $\gamma = 0.203$, $h = 1.0$ and $\gamma = 0.7$, $h = 0.3$ exhibits the reversal of EA ordering:  the state, $\gamma = 0.203$, $h = 1.0$ initially has less EA compared to the state, $\gamma = 0.7$, $h = 0.3$. But at late times, the order of EA is reversed. This indicates QME in the short-range interaction regime.

{\color{black}Thus, the QME occurs in all three interaction ranges, short, quasi-long and long-range regimes, for two suitably chosen sets of ground state parameters, whereas, for the tilted initial states with $\theta_1+\theta_2<\pi$, the phenomenon is completely absent in the quasi-long and the long-range regimes for the tilted initial states, as reported above. The analysis implies that the QME does not solely depend on $\alpha$ fixing the interaction range of the driving Hamiltonian, but also on the cooperative acts between the initial-state correlations and the post-quench dynamics. In the following, we understand this by employing the quasi-particle description.}

\begin{figure*}[t!]
    \centering
    \includegraphics[scale = 0.44]{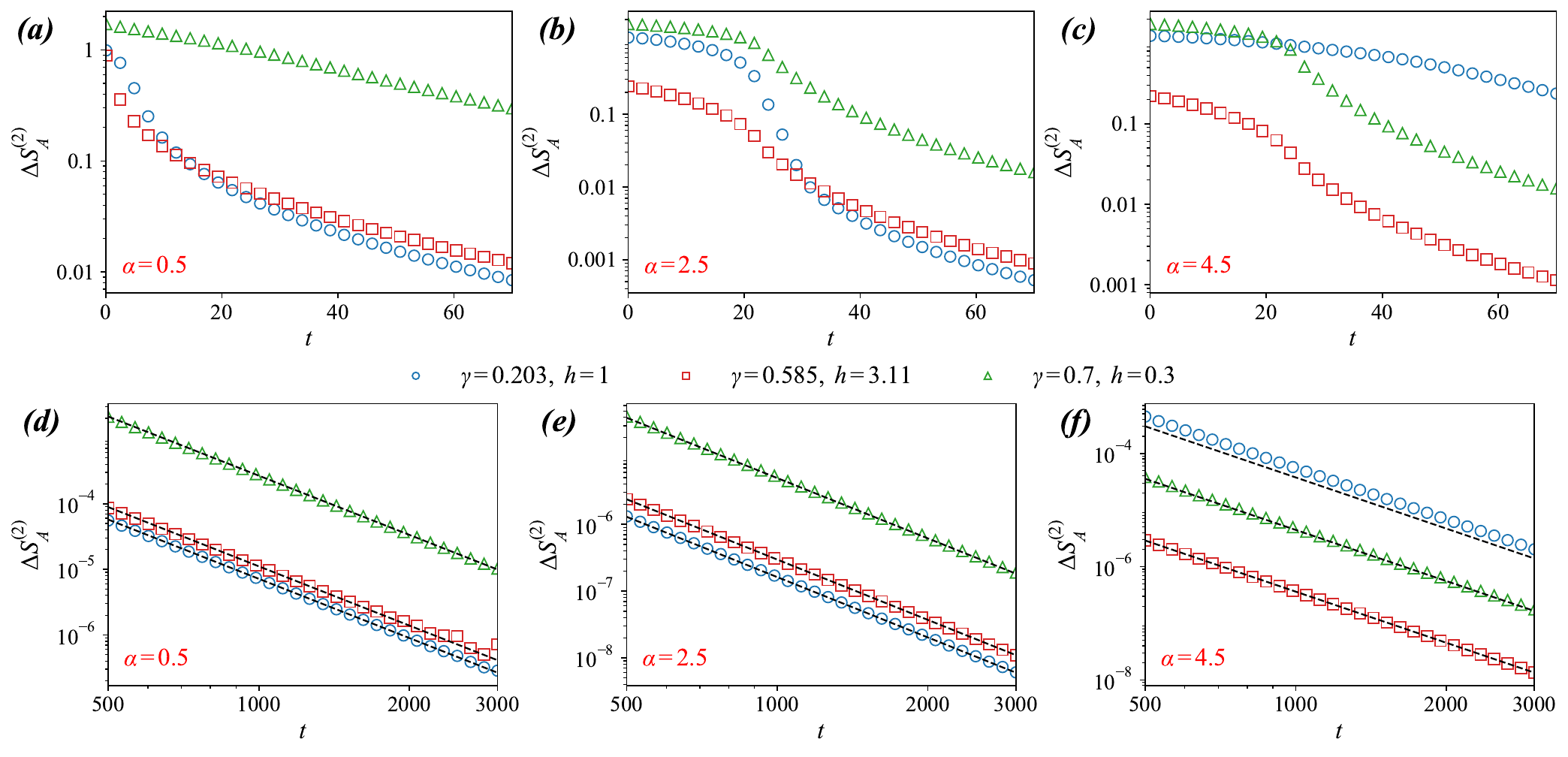}
    \caption{$\Delta S_A^{(2)}$ after a quench from the ground state of Eq.~(\ref{Hamiltonian}) for different $(h, \gamma)$ pairs, for $\alpha=0.5$, $2.5$ and $4.5$
(left to right). The top row (a)--(c) shows early-time behavior, while the bottom row (d)--(f) shows late-time behavior of EA. The dashed line in the bottom row is EA as calculated using the quasi-particle picture in the large-time limit. For every value of $\alpha$, there is QME among states. For $\alpha=0.5,2.5$, QME is shown among the states $(\gamma=0.203,h=1.0)$, represented by the blue circles, and $(\gamma=0.585,h=3.11)$, represented by the red squares. However, for $\alpha=4.5$, QME is shown among the states $(\gamma=0.203,h=1.0)$, represented by the blue circles, and $(\gamma=0.7,h=0.3)$, represented by the green triangles. Here, the subsystem size is taken to be $ l=50$.}
  \label{fig:Ent_asymmetry}
\end{figure*} 

\section{Quasi-particle interpretation} 
\label{quasi-particle}

The exact results above demonstrating QME or its absence for different values of the long-range interaction parameter $\alpha$ can be interpreted within the quasi-particle picture. This was first proposed to compute entanglement dynamics after a global quench in \cite{Calabrese_2005,Alba-Calabrese_2017} and has since been used to describe QME in short-range integrable systems \cite{Murciano2024,Ares2025,ares2023entanglement}. {\color{black}  Here we
extend it to the string-dressed long-range chain and use it to derive the conditions under which the QME occurs and to find its late-time asymptotic behavior.}

In the quasi-particle framework, 
each Cooper pair $(k,-k)$ of the initial state are the post-quench
quasi-particles that move apart with velocities $\pm V_{XX}(k)$. A pair contributes to the symmetry breaking only when both of its members are present inside the subsystem. At time
$t$, the pairs contributing to this are given by,
\begin{equation}
x_k(z) = \Big[1-\min\!\big(2\zeta|V_{xx}(k)|,1\big)\Big],
\label{xk}
\end{equation}
which is the fraction of the pairs created inside, where $z = t/l$ and $l$ is the subsystem size. Condition (ii) of Eq.~\eqref{QME_def} is satisfied at late times, when the asymmetry
is small. The late-time behavior of EA, $\Delta S_A^{(2)}(t)$ can be expressed as (see Appendix.~\ref{Quasi_P_Picture}),
\begin{equation}
\label{eq:quasi_particle_ent_asymmetry}
\Delta S_A^{(2)}(t) \simeq  2 l b(z,\alpha,\bar{\lambda}), 
\end{equation}
where
\begin{equation}
b(z,\alpha,\bar{\lambda}) = \int_{-\pi}^{\pi}\frac{dk}{4\pi}\, x_k(z)~\mathcal{G}_k(\alpha,\bar{\lambda}),
\end{equation}
with
\begin{equation}
\mathcal{G}_k(\alpha,\bar{\lambda})=-\ln\left(\frac{1 + \sqrt{1 - \sin^2 \Delta_k}}{2}\right),
\end{equation}
where $\bar{\lambda}$ denotes the parameters controlling asymmetry of the initial states, which are $\theta$ for the tilted state and $\{h,\gamma\}$ for the ground states.
Thus in the large-time regime, this condition for QME translates into a comparison between the corresponding $b$-functions, and hence, the emergence of QME at long times can be directly understood from the behavior of the function $b(\alpha,\bar{\lambda})$.  \\

Within the quasiparticle picture, the entanglement dynamics at large times is dominated by the slowest quasiparticle modes, since faster modes contribute at earlier times and saturate quickly. The behavior of $V_{xx}(k)$ depends crucially on the interaction exponent $\alpha$. The velocity diverges at $k=0$ for $\alpha<2$ in the thermodynamic limit, but assumes finite values with an interaction cut-off, as discussed above (see Fig.~\ref{group-velocity}), and there the slow modes originate from momentum regions near $k=\pm\pi$. In contrast, for larger values of $\alpha$ ($\alpha>2$), this divergence is absent, and the
velocity approaches zero near both $k=0$ and $k=\pm\pi$. Hence, slow modes emerges near $k=\pm 0$, along with near $k=\pm\pi$ (see Sec.~\ref{Velocity_asymptotics}). Consequently, in the large-$z$ limit, the dominant contribution to $b(z,\theta)$ arises from small neighbourhoods around these points.

This allows us to approximate  as,
\begin{equation}
\begin{split}
b(z,\alpha,\bar\lambda)\simeq{}&\Biggl[\int_{\pi-|k|<k_\pi^*(z)}
+\mathcal{S}(\alpha-2)\int_{|k|<k_0^*(z)}\Biggr]\\
&\times\frac{dk}{4\pi}\,x_k(z)\,\mathcal{G}_k(\alpha,\bar\lambda),
\end{split}
\label{eq:bapprox-1}
\end{equation}
where $\mathcal{S}$ is the Heaviside step function.
Here, $k_0^*(z)$ and $k_{\pm\pi}^*(z)$ define the momentum windows around $k=0$ and $k=\pm\pi$, respectively, determined by the condition $|2V_{xx}(k)| < z^{-1}$ in the large-$z$ limit. 

While the details of the calculations are provided in Appendix.~\ref{EA_infty}, here we provide the final late-time asymptotic expressions. The window at $k=\pm \pi$ is common throughout for any value of $\alpha$ and is also the simplest. It turns out, in units of $J=1$, 
\begin{equation}
\Delta S_A^{(2)}(t)\simeq\frac{l^{4}}{384\pi\,t^{3}}\,
\frac{c_\pi^{2}}{(\eta(\alpha-2))^3},
\label{late_pi}
\end{equation}
where the explicit form of $c_\pi$ depends upon the initial state parameters $c_\pi=c_\pi(\bar{\lambda})$ (see in Appendix.~\ref{EA_infty}). 
The window at $k=0$ depends on how the velocity behaves in different $\alpha$ regimes. The accountable contributions from the slow modes appear only if $\alpha>3$. Altogether, the leading late-time asymmetry is
\begin{equation}
\Delta S_A^{(2)}(t)\simeq\frac{l^{4}}{384\pi t^{3}}
\biggl[\frac{c_\pi^{2}}{(\eta(\alpha-2))^3}
+\mathcal{S}(\alpha-3)\frac{c_0^{2}}{(\zeta(\alpha-2))^3}\biggr],
\label{late_total}
\end{equation}
where $c_0=c_0(\bar{\lambda})$ (see in Appendix.~\ref{EA_infty}). 
The case of $\alpha=3$ is marginal. 

Depending on which pairs are slow, Eq.~(\ref{late_total}) distinguishes interaction ranges via the second term. The initial state enters only through $c_\pi$ and $c_0$. For $\alpha<3$, the late-time ordering of two states is fixed by the $c_\pi$-term alone, whereas for $\alpha>3$, the $c_0$-term also contributes. $c_\pi$ and $c_0$ depend distinctly on the parameters of the two classes of the initial states introduced in Sec.~\ref{QME}.

\begin{figure}[t]
    \centering
    \includegraphics[width=\columnwidth]{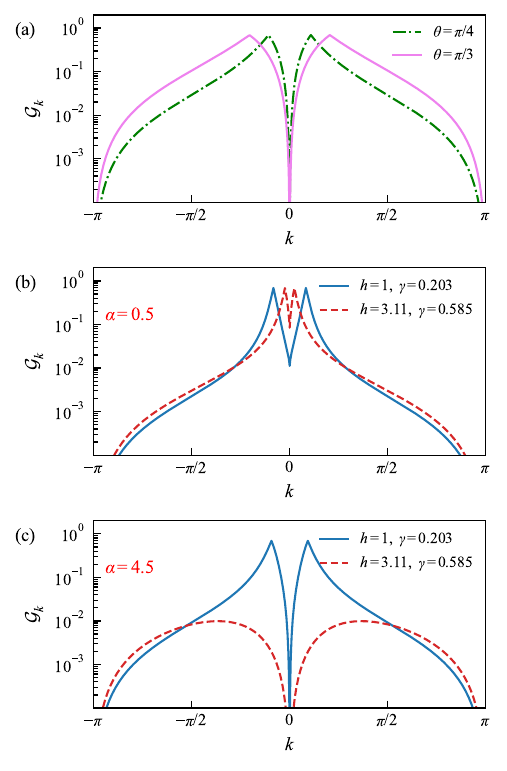}
    \caption{The plots (a),(b) and (c) show the quasiparticle density defined as $\mathcal{G}_k$ for different initial states and interaction range. Fig.(a) shows the quasiparticle density for tilted ferromagnetic states parametrized with representative tilt angles, $\theta=\pi/4$ and $\theta=\pi/3$. For this class of states, $\mathcal{G}_k$ does not depend on the interaction-range parameter $\alpha$. Figs.~(b) and (c) show the quasiparticle density for the ground states parametrized by $h,\gamma=(1,0.203) \text{ and } h,\gamma=(3.11,0.585)$ for $\alpha=0.5 \text{ and } \alpha=4.5$ respectively. The initial and late-time behaviour of EA for these representative initial states is displayed in the Fig.~\ref{fig:QME_tilted_st} and Fig.~\ref{fig:Ent_asymmetry}}. 
    \label{fig: QP_Density}
\end{figure}
\subsection{Tilted states}
\label{Tilted state QPP}
For the tilted states $c_0$ and $c_{\pi}$ are independent of $\alpha$. Their explicit forms are given by,
\begin{equation}
    c_{\pi} = \tan^{2}\frac{\theta}{2},
    \quad  c_0 = \cot^{2}\frac{\theta}{2}.
    \label{c_tilted} 
\end{equation}
Thus, the slow pairs near $k=\pm\pi$ contributing to the asymmetry 
grows with the tilt angle, whereas 
the contributions of the slow pairs near $k=0$ decrease.

For $\alpha < 3$, only the first term in Eq.~(\ref{late_total}) plays a part in, suggesting
\begin{equation}
\Delta S_A^{(2)}(t,\theta)|_{\alpha<3}\simeq\frac{l^{4}}{384\pi t^{3}}
\frac{\tan^4(\theta/2)}{(\eta(\alpha-2))^3}.
\label{late_tilited-long}
\end{equation}
Hence, at late times, EA increases monotonically over the whole range $0<\theta<\pi$. For a given pair of initial states with $\theta_1<\theta_2$, QME occurs when $\Delta S^{(2)}_A(\theta_1,t=0)>\Delta S^{(2)}_A(\theta_2,t=0)$. As the initial asymmetry is proportional to $\sin^2(\theta)$ (see Eq.~\ref{Asymmetry_Scaling_tilt}), the QME condition for the pair $(\theta_1,\theta_2)$ translates into $\sin^2(\theta_1)-\sin^2(\theta_2)>0$ which is only satisfied when $\theta_1+\theta_2>\pi$. For $\theta_1+\theta_2=\pi$, the initial EA is the same for both states. Such a pair does not display QME.  Noticeably, the window at $k=0$, which is completely absent for $\alpha<2$, weakly contributes for $2<\alpha<3$ in late times, such that $\Delta S_A^{(2)}$ has an additional  term $\propto c_0^{2}\,t^{-3/(\alpha-2)}$, which becomes sub-leading to Eq.~(\ref{late_tilited-long}) as $t$ grows. Figs.~\ref{fig:QME_tilted_st}(d) and \ref{fig:QME_tilted_st}(e) show exemplary cases for
$\alpha=0.5$ and $2.5$: the curves decay as $t^{-3}$ with the amplitudes of Eq.~(\ref{late_tilited-long}) (dashed lines), which agree quite well with the exact result, and their late-time ordering is the initial one. Thus it persists up to
$\alpha=3$, because the string dressing moves the slow pairs to $k=\pm\pi$, where
the more tilt remains dominant in the asymmetry during dynamics.

For $\alpha>3$ both terms of Eq.~(\ref{late_total}) contribute:
\begin{equation}
\Delta S_A^{(2)}(t)\simeq\frac{l^{4}}{384\pi t^{3}} f(\alpha,\theta),
\label{late_shortrange}
\end{equation}
where
\begin{equation}
f(\alpha,\theta)=\frac{\tan^{4}(\theta/2)}{(\eta(\alpha-2))^{3}}
+\frac{\cot^{4}(\theta/2)}{(\zeta(\alpha-2))^{3}}.
\label{eq:F_theta}
\end{equation}
The two terms above have opposite dependencies on $\theta$ and provide a leading-order explanation for the occurrence of the QME when $\alpha>3$. {\color{black} For a pair of tilted initial states with tilt angles $\theta_1<\theta_2$, the initial EA order is $\Delta S^{(2)}_A(\theta_1,t=0)>\Delta S^{(2)}_A(\theta_2,t)$ only when $\theta_1+\theta_2>\pi$. For this class of states, the order of EA reverses at late times, $\Delta S^{(2)}_A(\theta_1,t \to \infty)<\Delta S^{(2)}_A(\theta_2,t \to \infty)$, therefore showing QME. Contrary to this, when $\theta_1+\theta_2<\pi$, the initial EA order is $\Delta S^{(2)}_A(\theta_1)< \Delta S^{(2)}_A(\theta_2)$. Hence, the condition for QME is $f(\alpha,\theta_1)>f(\alpha,\theta_2)$
\begin{equation}
\begin{split}
    \tan^2\left(\frac{\theta_1}{2}\right)\tan^2\left(\frac{\theta_2}{2}\right)<\left(\frac{\eta(\alpha-2)}{\zeta(\alpha-2)}\right)^{3/2},
    \label{QME-criteria-1a}
\end{split}
\end{equation}
or alternatively, using $\eta(s)=(1-2^{1-s})\zeta(s)$, as
\begin{equation}
     \tan^2\left(\frac{\theta_1}{2}\right)\tan^2\left(\frac{\theta_2}{2}\right)<(1-2^{3-\alpha})^{3/2}.
     \label{QME-condition-1p}
\end{equation}
 
The function $f(\alpha,\theta)$ has a minimum at 
\begin{equation}
\tan^{2}\left(\frac{\theta^{*}}{2}\right)=
\left[\frac{\eta(\alpha-2)}{\zeta(\alpha-2)}\right]^{3/4},
\label{eq:theta_star}
\end{equation}
so that every pair with $\theta_1<\theta_2\le\theta^{*}(\alpha)$ displays the QME. In the large $\alpha$ limit, $\alpha \to \infty$, $\theta^* \to \pi/2$, while for $\alpha \to 3$, $\theta^* \to 0$.

Figs.~\ref{fig:QME_tilted_st}(c) and \ref{fig:QME_tilted_st}(f) show an exemplary cases for
$\alpha=4.5$. It can be clearly seen that the late-time ordering of the EA for a pair of slates with $\theta_1=\pi/4$ and $\theta_2=\pi/3$ is reversed when compared with the initial ordering, and thus displaying the QME. The analytical result presented in Eq.~(\ref{late_shortrange}) is compared with the exact results, and they agree. The pair of angles considered here satisfies the leading condition for QME in Eq.~(\ref{QME-condition-1p}). 

The presence or absence of QME can be understood within the quasiparticle picture by using the Cooper-pair density $\mathcal{G}_k$(Fig. \ref{fig: QP_Density}). We are taking the representative example of the states with $\theta_1=\pi/4$ and $\theta_2=\pi/3$. {\color{black} At late times, only the slowest quasiparticles are shared between the subsystem $A$ and the rest of the system, $\bar{A}$}, so the late-time behavior of EA is determined by the weight of $\mathcal{G}_k$ near the zeros of quasiparticle velocity, $|V_{XX}(k)|$. For tilted ferromagnetic states, $\mathcal{G}_k$ depends only on tilt-angle $\theta$ (Fig.~\ref{fig: QP_Density}(a)). Near $k=0$, $\mathcal{G}_k \propto \cot^4(\theta/2)$, having larger value for smaller tilt-angle $\theta$ and near $k=\pm\pi$, $\mathcal{G}_k \propto \tan^4(\theta/2)$, which increases with tilt-angle $\theta$. For $\alpha<2$, $|V_{XX}(k)|$ diverges as $k \xrightarrow{} 0$, so the quasiparticle pairs near $k=0$ escape from the subsystem instantly. At late times, only the pairs around $k=\pm\pi$ survive and contribute to the behavior of EA. For $\alpha\in(2,3)$, near $k=0$, $|V_{XX}(k)| \sim |k|^{\alpha-2}$, so this channel contributes in late time EA as $\sim t^{-3/(\alpha-2)}$ that is subleading with respect to $\sim t^{-3}$ contribution coming from $k=\pm\pi$.  In both of the cases, the state with the larger tilt-angle $\theta$ retains more slow-pairs. For $\theta_1+\theta_2<\pi$, the state with the higher tilt angle is also the initially more asymmetric
state. Hence, the initial ordering of EA remains preserved at late times, resulting in no QME in this interaction regime (Fig.~\ref{fig:QME_tilted_st}(d),(e))}. In Fig.~\ref{fig: QP_Density}(a), it is clearly shown that quasiparticle density is larger for $\theta=\frac{\pi}{3}$ (pink-solid line) than it is for $\theta=\frac{\pi}{4}$ (green-dash-dotted line) around $k=\pm\pi$. Hence, EA in late time, for $\theta=\frac{\pi}{3}$ is more than it is for $\theta=\frac{\pi}{4}$ in $\alpha<3$ regime. Although in the short-range regime, both $k=0$ and $k=\pm\pi$ channels contribute to late-time EA as $\sim t^{-3}$. In the representative case for $\alpha=4.5$ with tilted states parameterized by $\theta_1=\pi/4\text{ and } \theta_2=\pi/3$, the $k=0$ channel dominates through its $\cot^4(\theta/2)$ weight, so the less tilted state $\theta_1=\pi/4$ keeps more slow pairs than $\theta_2=\pi/3$ (Fig.~\ref{fig: QP_Density}(a)). The more asymmetric state therefore restores symmetry faster, and the curves cross 
(Figs.~\ref{fig:QME_tilted_st}(c),(f)).

{\color{black} The tilted states thus display QME in all the interaction regimes, although it is absent in the long-range regime when $\theta_1+\theta_2<\pi$.
 The latter behaviour is in contrast with the results reported in the long-range systems without the string dressing \cite{Yamashika2026}, where the QME occurs in the long-range regime with a pair of states satisfying $\theta_1+\theta_2<\pi$.}

\subsection{Ground states}
Unlike the tilted states, $c_\pi$ and $c_0$ for the ground state depends upon $\alpha$:
\begin{equation}
c_\pi=\frac{\gamma~(\eta(\alpha-1))}{(h+\eta(\alpha))}, \quad c_0=\frac{\gamma~(\zeta(\alpha-1))}{|h-\zeta(\alpha)|}.
\label{cpi-ground}
\end{equation}
Similar to the tilted states, we discuss the regimes $\alpha<3$ and $\alpha>3$ separately.

For $\alpha<3$, the initial asymmetry increases monotonically with $g(\alpha,\gamma,h)$ (Eq.~(\ref{eq_g})). For $\alpha<3$, the late-time asymmetry is proportional to $c_\pi$ only. Two ground states therefore display the QME for $\alpha<3$ when

\begin{align}
g(\alpha,\gamma_1,h_1)>&g(\alpha,\gamma_2,h_2),\nonumber\\
\frac{\gamma_1}{h_1+\eta(\alpha)}<&\frac{\gamma_2}{h_2+\eta(\alpha)}.
\label{gs_QME_q1}
\end{align}
We find that Eq.~(\ref{gs_QME_q1}) can be satisfied by suitably chosen pairs in every interaction range when $\alpha<3$, and thus resulting in QME.

In the Figs.~\ref{fig:Ent_asymmetry}(a) and \ref{fig:Ent_asymmetry}(b), we present representative cases for $\alpha=0.5$ and $\alpha=2.5$ respectively. The pairs $(\gamma=0.203,h=1.0)$, and $(\gamma=0.585,h=3.11)$,  exhibiting QME for both cases satisfies Eq.~(\ref{gs_QME_q1}). The dashed line showing the late-time expression in Eq.~(\ref{late_total}) with $c_\pi$ given in Eq.~(\ref{cpi-ground}) agrees well with the exact results for all the initial states examined in this work (see Fig.~\ref{fig:Ent_asymmetry}).   

For $\alpha>3$, the QME condition is modified. Both the terms in Eq.~(\ref{late_total}) now contribute; modes near $k=\pm \pi$ and $k=0$ come into play. Interestingly, the
denominator of $c_0$ brings additional implications to the QME. A state whose field lies
close to the critical field $h_c=\zeta(\alpha)$ carries a large weight on the slow
pairs near $k=0$, whatever its initial asymmetry. Consequently, its asymmetry is restored slowly. When such a state is paired with a state that is more asymmetric
initially but has little weight on the slow modes, e.g., away from criticality, eventual occurrence of the QME is expected. We show this at $\alpha=4.5$ (Figs.~\ref{fig:Ent_asymmetry}) for a state $(0.203,1.0)$ with near-critical field, with criticality at $h_c = \zeta(4.5) \sim 1.055$, and compare it with $(0.7,0.3)$ and 
$(0.585,3.11)$. The most asymmetric initial state $(0.7,0.3)$  therefore restores the symmetry faster than the near-critical state. The pair that displays the QME for $\alpha<3$ no longer does so in this case. Exact results and dashed lines from Eqs.~(\ref{late_pi},\,\ref{late_total}) in Fig.~\ref{fig:Ent_asymmetry} agree.

We can similarly explain the occurrence of QME using the quasiparticle density $\mathcal{G}_k$ in different $\alpha$ regimes for the ground state. Unlike the tilted ferromagnetic state, $\mathcal{G}_k$ depends on the interaction parameter $\alpha$ (see in Figs.~\ref{fig: QP_Density}(b),(c)). We discuss the behavior of EA at late time using the quasiparticle density $\mathcal{G}_k$ for representative initial states with $(h,\gamma)=(1.0,0.203)$ and $(h,\gamma)=(3.11,0.585)$ for the decay exponents $\alpha=0.5$ and $\alpha=4.5$. In long-range interaction regime, for $\alpha=0.5$, $|V_{XX}(k)|$ diverges at $k=0$ (Fig.~\ref{group-velocity}(a)). As the quasiparticle pairs around $k=0$ leave the subsystem instantly, they do not contribute to the late-time behaviour of EA. Only the pairs near $k=\pm\pi$ contribute. Near $k=\pm\pi$, the quasiparticle density scales as $\mathcal{G}_k \propto c_{\pi}^2$. For $\alpha=0.5$, this coefficient takes the values $c_{\pi}^2 \approx 2.31\times10^{-3}$ for $(h,\gamma)=(1.0,0.203)$ and $c_{\pi}^2 \approx 3.58\times10^{-3}$ for $(h,\gamma)=(3.11,0.585)$. Therefore, around $\pm \pi$, $(h,\gamma)=(1.0,0.203)$ state has less density of quasiparticles than $(h,\gamma)=(3.11,0.585)$ state (Fig.~\ref{fig: QP_Density}(b)). This results in a reversal of the order of EA at late times, and hence shows QME (Fig.~\ref{fig:Ent_asymmetry}(a),(d)). For $\alpha=4.5$, the late-time contribution in EA is due to the quasiparticle pairs near both $k=0$ and $k=\pm \pi$ as $|V_{XX}(k)| \xrightarrow{} 0$ (see Fig.~\ref{group-velocity}). Near $k=0$, quasiparticle density scales as $\mathcal{G}_k \propto c_0^2$. For $\alpha=4.5$, the coefficients take the values $c_0^2 \approx 17.48,c_{\pm\pi}^2 \approx 9.22 \times10^{-3}$ for the state $(h,\gamma)=(1.0,0.203)$ and  $c_0^2 \approx 0.103,c_{\pm\pi}^2 \approx 1.78 \times10^{-2}$ for the state $(h,\gamma)=(3.11,0.585)$. Although there is a competition between contributions from $k=0$ and $k=\pm\pi$, it is clearly evident from Fig.~\ref{fig: QP_Density}(c) that the dominating contribution of slow-mode density comes from $k=0$ neighbourhood, which in turn determines the order of EA for the given states. Hence, the state $(h,\gamma)=(1.0,0.203)$ is more asymmetric than the state $(h,\gamma)=(3.11,0.585)$ at large time, retaining the initial order of EA. So there is no QME between this pair of states for $\alpha=4.5$  (Figs.~\ref{fig:Ent_asymmetry}(c),(f)).

\section{Discussion} 
In this work, we consider a string-dressed long-range quantum spin chain with a power-law interaction, with $\alpha$ as the exponent controlling the interaction profile, and study EA and QME. In the fermionic representation, the model is quadratic and hence integrable--the system can be tracked exactly for all interaction ranges of $\alpha$. We considered two classes of initial states with broken $U(1)$-symmetry, the tilted ferromagnetic states and the ground states of the anisotropic chain, and consequently performed sudden quenching governed by the $U(1) $- symmetry-preserving Hamiltonian, for which we set $h=\gamma=0$.

We first investigate EA of the system in equilibrium. EA has a logarithmic dependence on the subsystem size, along with a logarithmic contribution stemming from charge fluctuations of the subsystem in all the interaction ranges. EA becomes non-analytic at the gapless point, serving as an indicator of the QPT, including the range of $\alpha<1$, where the critical field is finite only upon the imposition of a finite-distance interaction cut-off, $N_c$, and it grows as $N_c^{1-\alpha}$.

Next, we study the out-of-equilibrium system and investigate QME. There, the two classes often behave differently, and the occurrence of the QME is conditioned upon the initial state parameters and $\alpha$ deciding the interaction range. For a pair of tilted states with $\theta_1+\theta_2>\pi$, the QME occurs for every $\alpha$, whereas for
$\theta_1+\theta_2<\pi$, it is absent for every $\alpha<3$. The underlying conditions are followed from the Eqs.~(\ref{QME-criteria-1a}-\ref{QME-condition-1p}). The system exhibits QME for all pairs of initial states below a threshold angle, $\theta^*(\alpha)$, for $\alpha>3$, which approaches the nearest-neighbour value $\theta^*(\alpha) \to \pi/2$ when $\alpha \to \infty$.  For ground states as the initial state, by contrast, the QME occurs for a suitable pair in all the interaction regimes.

In all the interaction regimes, the QME finds a physical explanation via the quasiparticle mechanism discussed in Sec.~{\ref{quasi-particle}}. Within the quasiparticle picture, the initial states are the generators of the quasiparticle pairs that move apart with group velocities determined by the dispersion. The symmetry is eventually restored as the asymmetry-carrying quasiparticles leave the subsystems, and thus the late-time asymmetry and the eventual fate of the QME are decided by the slowest modes.  The interaction range decides which pairs are slow, and the initial state decides how much asymmetry they carry. In all interaction ranges, the neighborhood of $k=\pm \pi$ hosts the slow modes, whereas in the short-range regime slow modes are also pinned near $k=0$.  The quasi-long-range regime, $2<\alpha<3$, accommodates slow modes near $k=0$, but there the velocity vanishes non-analytically in this range. The velocity vanishes linearly
at $k=\pm\pi$ for every $\alpha$, and at $k=0$ for $\alpha>3$, whereas it diverges at $k=0$ when $\alpha<2$. These modes are the fastest ones and leave the subsystems immediately, and hence do not play any role in the late-time ordering. As a result, the asymptotic law obtained with the thermodynamic velocity agrees with the exact results at large cutoff $N_c$ for all values of $\alpha$ studied in this work. The quasiparticle picture therefore applies throughout the long-range regime. 

Finally, a few aspects deserve highlighting. First, the boundary at $\alpha=3$ for the tilted initial states reflects competition between slow Cooper pairs near $k=\pm\pi$ and $k=0$, carrying tilt-dependent weights via $c_\pi$ and $c_0$, which, respectively, increases or decreases with increasing tilt angle. While the former always contribute, the latter enter at the same $t^{-3}$ order only for $\alpha>3$, allowing a reversal of the late-time asymmetry ordering for $\theta<\theta^*(\alpha)$. For $\alpha<3$, the QME is decided by the initial ordering alone and occurs if and only if $\theta_1+\theta_2>\pi$. This contrasts with long-range spin chains without the string dressing, where the QME arises from magnetization fluctuations that restore symmetry at long range ~\cite{Yamashika2026} and the QME has been reported to be present for all considered pairs with $\theta_1+\theta_2<\pi$ in the long-range regime
$0\le\alpha\le1$, and thus indicates the governing mechanism of the dressed chains to be distinct from the bare long-range systems. Secondly, for $\alpha>3$, proximity to the critical field provides an alternative means by placing a large pairing amplitude on slow modes and making the near-critical state the slower one to restore the symmetry. This aspect has been reported in studies involving the nearest-neighbor XY spin chain as well \cite{Murciano2024}.

The QME has been demonstrated in the ion trap set-up, realising an undressed long-range spin chain. The dressed chain is a long-range Kitaev chain, and both kinds of the initial states considered in this work are Gaussian in nature. Gaussian states are preparable in digital processors, while the quench via the driving Hamiltonian can be compiled into a fermionic-swap network \cite{Jiang2018,Kivlichan2018}. The role of the string dressing can then be realized by comparing it with the undressed chain at the equal-interaction point in this set-up.

\section{Appendix}
\appendix

\subsection{Charged Moments and Entanglement Asymmetry}
\label{Charged Moments}
To calculate $\Delta S^{(n)}_A$, we need to first calculate charged moments, $Z_n(\boldsymbol{\tilde\alpha})$ and put it back to Eq.~(\ref{eq:entanglement_asymmetry_charged}). Charged moments are defined as, 

\begin{equation}
\label{eq:charged_moments}
Z_n(\boldsymbol{\tilde{\alpha}})
=
\mathrm{Tr}\Bigg[\prod^n_{j=1} \rho_A e^{i(\alpha_j-\alpha_{j+1})Q_A}\Bigg],
\end{equation}

with $\boldsymbol{\tilde\alpha}= (\alpha_1,\alpha_2,...,\alpha_n)$. Here $\rho_A$ is the reduced density matrix of subsystem $A$, and $Q_A$ denotes the conserved $U(1)$ charge operator restricted to subsystem $A$. Using Jordan-Wigner transformation, $Q_A$ can be written as $Q_A=\sum_{j\in A} c_j^{\dagger}c_j$ and $e^{i\alpha Q_A}$ turns out to be gaussian. The reduced density matrix, $\rho_A$, for both the ground state and the tilted ferromagnetic state is Gaussian. Using Wick's theorem together with the Gaussian structure of the state, the charged moments can be expressed entirely in terms of the correlation matrix of the subsystem. The two point correlation matrix $\Gamma$ of the subsystem with size $l$ is,
\begin{equation}
\begin{split}
    \Gamma &= 2\,\langle \Psi \Psi^{\dagger} \rangle - \mathbb{I}_{2l}
           = 2\,\mathrm{Tr}\!\left[\rho_A\, \Psi \Psi^{\dagger}\right] - \mathbb{I}_{2l},\\
    \qquad
    \Psi &= \left( c_1,\, c_1^{\dagger},\, c_2,\, c_2^{\dagger},\, \dots,\, c_l,\, c_l^{\dagger} \right)^{T}.
\end{split}
\end{equation}
The charged moments, $Z(\boldsymbol{\tilde\alpha)}$ can be expressed in term of $\Gamma$ as \cite{Murciano2024, SciPostPhys.15.3.089}, 
\begin{equation}
    Z_n(\boldsymbol{\tilde\alpha}) = \sqrt{\det\left[\left(\frac{I-\Gamma}{2}\right)^{n}\left(I+\prod_{j=1}^{n} W_j\right)\right]},
    \label{eq:Zn}
\end{equation}
with $W_j = (I+\Gamma)(I-\Gamma)^{-1} e^{i\alpha_{j,j+1} n_A}$ and $n_A$ is a diagonal matrix with
$(n_A)_{2j,2j} = -1$, $(n_A)_{2j-1,2j-1} = 1$, $j = 1, \ldots, l$.
Eq.~(\ref{eq:Zn}) allows to exactly compute numerically $\Delta S_A^{(n)}$ using Eq.~(\ref{eq:entanglement_asymmetry_charged}). Here we are calculating only $\Delta S_A^{(2)}$ so we only need $Z_2(\tilde\alpha)$,
{\color{black}\begin{equation}
\label{eq:Zn_det}
Z_2(\tilde{\alpha})
=
\det
\left(
\frac{
I+\Gamma_{\tilde{\alpha}}\Gamma_{-\tilde{\alpha}}
}{2}
\right)^{1/2}.
\end{equation}}
with $\tilde\alpha=\alpha_1-\alpha_2$ and $\Gamma_{\tilde\alpha}=\Gamma e^{i\tilde\alpha n_A}$. Due to translation invariance of the system, the momentum space correlation matrix of the whole system is block diagonal, with each $(k,-k)$ pair decoupled from the rest. The $2 \times2$ blocks are,

\begin{equation}
G(k) =
\begin{pmatrix}
\cos\Delta_k & -i\sin\Delta_k \\
i\sin\Delta_k & -\cos\Delta_k
\end{pmatrix},
\label{G(k)}
\end{equation}
where $\Delta_k$ is the Bogoliubov angle of the initial state. $e^{i\tilde\alpha n_A}$ is block-diagonal with $2 \times 2$ blocks $e^{i\tilde\alpha\sigma_{z}}$. The matrix elements of $\Gamma_{\tilde\alpha}$ is $(\Gamma_{\tilde{\alpha}})_{jj'}=\Gamma_{jj'}e^{i\tilde\alpha\sigma_z}$ which can also be written as,
\begin{equation}
    (\Gamma_{\tilde{\alpha}})_{jj'}=\int^{\pi}_{-\pi} \frac{dk}{2\pi} G_{\tilde\alpha}(k) e^{-ik(j-j')},\quad j,j'\in\{1,2,...,l\}
\end{equation}
here, 
\begin{equation}
G_{\tilde\alpha}(k) =
\begin{pmatrix}
e^{i\tilde{\alpha}}\cos\Delta_k & -i e^{-i\tilde{\alpha}}\sin\Delta_k \\
i e^{i\tilde{\alpha}}\sin\Delta_k & -e^{-i\tilde{\alpha}}\cos\Delta_k
\end{pmatrix},
\end{equation}
Thus, we can calculate the correlation matrix of the subsystem with length $l$ from initial state parameters and, in turn, can calculate the charge moment $Z_2(\tilde \alpha)$ and EA,
\begin{equation}
    \Delta S^{(2)}_A=-\log\Big(\int^{\pi}_{-\pi} \frac{d\tilde\alpha}{2\pi} \frac{Z_2(\tilde \alpha)}{Z_2(0)}\Big).
    \label{Z2_int}
\end{equation}
\subsection{Scaling of Entanglement Asymmetry}
\label{Scaling of EA}
We can calculate the charged moment, $Z_2(\tilde\alpha)$, as a function of subsystem size $l$. The matrix $\Gamma_{\tilde\alpha}$ is block-Toeplitz. If $T_l[g]$ is a block-Toeplitz matrix of dimension $(ld) \times (ld)$ with $d \times d$ matrix $g$, then for large $l$\cite{SciPostPhys.15.3.089,Murciano2024}, 
\begin{equation}
    \det\left( I + \prod_{j=1}^{n} T_{l}[g_j] \right) \sim e^{l A},
    \label{Toeolitz scaling}
\end{equation}
where the coefficient $A$ is given by
\begin{equation}
    A = \int_{0}^{2\pi} \frac{\mathrm{d}k}{2\pi} \, \log \det\left[ I + \prod_{j=1}^{n} g_j(k) \right].
\end{equation}
Using this, we can write the charged moment $Z_2(\tilde \alpha)$ in large $l$ limit,
\begin{equation}
    Z_2(\tilde\alpha) \sim \frac{1}{2^l} \times \exp\Big(\frac{l}{2}\int^{2\pi}_0\frac{dk}{2\pi}\log(\det[I+G_{\tilde\alpha}(k) G_{-\tilde\alpha}(k)])\Big).
\end{equation}
Then this can be further simplified to
\begin{equation}
    Z_2(\tilde\alpha)=Z_2(0)\exp\Big(\frac{l}{2}\int^{2\pi}_0\frac{dk}{2\pi}\log(1-\sin^2(\tilde\alpha)\sin^2(\Delta_k)\Big).
    \label{Z2_ini}
\end{equation}
Hence, $\Delta S^{(2)}_A$ expression in large $l$ limit is given by, 
\begin{equation}
    \Delta S^{(2)}_A \simeq -\log \int_{-\pi}^{\pi} \frac{d\tilde\alpha}{2\pi}\, e^{\, l f(\tilde\alpha)},
    \label{analytic_EA_2}
\end{equation}
where
\begin{equation}
    f(\tilde\alpha) = \int_{0}^{2\pi} \frac{dk}{4\pi}\,
    \log\!\left(1 - \sin^2\tilde\alpha \, \sin^2\Delta_k\right).
\end{equation}
For large $l$, the integrand $e^{\,l f(\tilde\alpha)}$ is sharply peaked. Since
$0 \le 1 - \sin^2\tilde\alpha\,\sin^2\Delta_k \le 1$, we have $f(\tilde\alpha) \le 0$,
with equality only where $\sin\tilde\alpha = 0$, i.e.\ at $\tilde\alpha = 0$ and
$\tilde\alpha = \pm\pi$. Away from these points, the integrand has negligible weight.

Near $\tilde\alpha = 0$, using $\sin^2\tilde\alpha \simeq \tilde\alpha^2$ and
$\log(1-x) \simeq -x$,
\begin{equation}
    f(\tilde\alpha) \simeq -\tilde\alpha^2 \int_{0}^{2\pi} \frac{dk}{4\pi}\sin^2\Delta_k.
\end{equation}
Using $2\pi$-periodicity of $\Delta_k$ to write
$\int_{0}^{2\pi} \frac{dk}{2\pi}\sin^2\Delta_k = \int_{-\pi}^{\pi} \frac{dk}{2\pi}\sin^2\Delta_k=g(\alpha, \bar{\lambda})$. Here, $\bar\lambda$ is fixed by the initial state, and $\alpha$ is the interaction range parameter.
The integration turns out to be a Gaussian integral, as the integration limits can be extended to $\pm\infty$.
\begin{equation}
    \int \frac{d\tilde\alpha}{2\pi}\, e^{-l g \tilde\alpha^2/2}
    = \frac{1}{2\pi}\sqrt{\frac{2\pi}{l g}} .
\end{equation}
Similarly, adding the contribution from $\tilde\alpha = \pm\pi$ we get,
\begin{equation}
    \int_{-\pi}^{\pi} \frac{d\tilde\alpha}{2\pi}\, e^{\,l f(\tilde\alpha)}
    \simeq 2 \cdot \frac{1}{2\pi}\sqrt{\frac{2\pi}{l g}}
    = \sqrt{\frac{2}{\pi\, g(\gamma,h)\, l}} .
\end{equation}

Taking $-\log$ gives
\begin{equation}
    \Delta S^{(2)}_A 
    = \frac{1}{2}\log (\frac{\pi\, g(\alpha,\bar{\lambda})l}{2})
    + \mathcal{O}(l^{-1}),
\end{equation}
which is Eq.~(\ref{Asymmetry_Scaling}) with $g(\alpha, \bar \lambda)=g(\alpha,h,\gamma)$.

\subsection{Charged Moment $Z_2(\tilde\alpha)$ at late times}
\label{Asymptotic Z_2}
We have studied the time evolution of the system after a global quench from $1.$ ground state of the long-range XY chain (Eq.~(\ref{GS})) $2.$ tilted ferromagnetic state (Eq.~(\ref{tilted})), with post-quench Hamiltonian which is $\gamma=0$ and $h=0$ limit of Eq.\ref{Hamiltonian}. Since the post-quench Hamiltonian is also quadratic in terms of fermionic operators $c_j$, the time-evolved reduced density matrix of the subsystem $\rho_A(t)$ is also Gaussian. The post-quench Hamiltonian preserves the translation invariance of the system. Hence, we can calculate the time-dependent charge moments in terms of the time-evolved two-point correlation matrix of the subsystem of length $l$.

The time-dependent correlation matrix restricted to a subsystem of length $l$ can be expressed as
\begin{equation}
\Gamma_{jj'}(t)
=
\int_{-\pi}^{\pi}
\frac{dk}{2\pi}
\,e^{-ik(j-j')}
\,G(k,t),
\qquad
j,j'=1,\dots,l ,
\label{eq:gamma_app}
\end{equation}
where the $2 \times2$ momentum pair $(k,-k)$ block is given by,
\begin{equation}
G(k,t)=
\begin{pmatrix}
\cos\Delta_k &
-i e^{-2it\epsilon_{XX}(k)}\sin\Delta_k
\\
i e^{2it\epsilon_{XX}(k)}\sin\Delta_k &
-\cos\Delta_k
\end{pmatrix}.
\label{eq:symbol_app}
\end{equation}
Here, $\epsilon_{XX}(k)=-\sum_{r=1}^{N_c}\frac{\cos(kr)}{r^\alpha}$ denotes the single-particle dispersion relation of the post-quench Hamiltonian $H_{XX}$, while $\cos\Delta_k$ and $\sin\Delta_k$ are defined in the main text. We can calculate the time-dependent charge moment,
\begin{equation}
    Z_2(\tilde \alpha,t)=\Bigg[\det\Bigg(\frac{I+\Gamma_{\tilde\alpha}(t)\Gamma_{-\tilde\alpha}(t)}{2}\Bigg)\Bigg]^{\frac{1}{2}}
\end{equation}
here, 
\begin{equation}
(\Gamma_{\tilde\alpha})_{jj'}(t)
=
\int_{-\pi}^{\pi}
\frac{dk}{2\pi}
\,e^{-ik(j-j')}
\,G_{\tilde \alpha}(k,t),
\qquad
j,j'=1,\dots,l ,
\label{eq:gamma_app}
\end{equation}
with, 
\begin{equation}
G_{\tilde \alpha}(k,t)=
\begin{pmatrix}
e^{i\tilde\alpha}\cos\Delta_k &
-ie^{-i\tilde\alpha} e^{-2it\epsilon_{XX}(k)}\sin\Delta_k
\\
i e^{i\tilde\alpha}e^{2it\epsilon_{XX}(k)}\sin\Delta_k &
-e^{-i\tilde\alpha}\cos\Delta_k
\end{pmatrix}.
\label{eq: G_alpha(k,t)}
\end{equation}
In the $t \xrightarrow[]{}\infty$ limit, the off-diagonal terms go to zero.
\begin{equation}
G_{\tilde\alpha}(k,t\to\infty)=
\begin{pmatrix}
e^{i\tilde \alpha}\cos\Delta_k & 0\\
0 & -e^{-i\tilde\alpha}\cos\Delta_k
\end{pmatrix}.
\label{eq:G_alpha(k,infty)}
\end{equation}
For $n=2$, the stationary behavior of $Z_2(\tilde\alpha,t\xrightarrow[]{} \infty)$ can be determined by applying Eq.~(\ref{Toeolitz scaling}), 
\begin{equation}
\begin{aligned}
\log Z_2(\tilde{\alpha}, t\to\infty) &\sim{}\frac{l}{2} \int_{0}^{2\pi} \frac{dk}{2\pi} 
\log\Bigg(\\&\det\left[\frac{I + G_{\tilde{\alpha}}(k, t\to\infty)\, G_{-\tilde{\alpha}}(k, t\to\infty)}{2}\right]\Bigg),
\end{aligned}
\label{eq:Z2_alpha_infty}
\end{equation}
Calculating the determinant inside of the integral in Eq.~\eqref{eq:Z2_alpha_infty}, we obtain
\begin{equation}
\log Z_2(\tilde{\alpha}, t\to\infty) \sim l\int_{0}^{2\pi} \frac{dk}{2\pi}\,
\log\left(\frac{1+\cos^2\Delta_k}{2}\right),
\label{eq:Z2_infty_explicit}
\end{equation}
which is independent of $\tilde{\alpha}$. This result implies that $Z_2(\tilde\alpha,t \xrightarrow[]{} \infty) \sim Z_2(0,t \xrightarrow[]{} \infty)$. At this point, we have the expression for charged moments at both $t=0$ and $t \xrightarrow{} \infty$. At any intermediate time $t$, the time-dependent charged moment $Z_2(\tilde\alpha,t)$ can be calculated using the quasiparticle picture as described in the next section.

\subsection{Quasiparticle Picture and time evolution of Entanglement Asymmetry}
\label{Quasi_P_Picture}
The initial state has higher energy compared to the ground state of the post-quench Hamiltonian. Hence, the initial state has excitations in terms of quasiparticle pairs with momentum $(k,-k)$. The quasiparticle pairs are created at a spatial point and disperse with velocity $V_{XX}(k)$(see Eq.~(\ref{VXX})). At generic time $t$, the entanglement between subsystem $A$ and $\bar{A}$ is proportional to the total number of quasiparticles shared between $A$ and $\bar{A}$, which is given by the function $\min(2t|V_{XX}(k),l)$.
\par Following the quasiparticle framework, the charged moments acquire the scaling form \cite{Murciano2024}

\begin{equation}
Z_2(\tilde{\alpha},t)
=
Z_2(0,t)\,
e^{\,l\left[
A_2(\tilde{\alpha})
+
B_2(\tilde{\alpha},z)
\right]}.
\label{eq:Zn_scaling_app}
\end{equation}
The dynamical contribution $Z_2(0,t)$ is given by

\begin{equation}
\begin{split}
\log Z_2(0,t)={}&l\int_{0}^{2\pi}\frac{dk}{2\pi}\,\min\!\bigl(2z|V_{XX}(k)|,1\bigr)\\
&\times\log\frac{1+\cos^{2}\Delta_k}{2},
\end{split}
\label{eq:Zn0_app}
\end{equation}
where $V_{XX}(k)$ denotes the quasiparticle group velocity. The stationary contribution $A_2(\tilde\alpha)$ is given as, 
\begin{equation}
    A_2(\tilde \alpha)=\int^{2\pi}_0 \frac{dk}{4\pi}\log(1-\sin^2(\tilde \alpha)\sin^2(\Delta_k))
\end{equation}
while the ballistic correction term reads
\begin{equation}
\begin{split}
B_2(\tilde{\alpha},z)
= -\int_{0}^{2\pi}\frac{dk}{4\pi}\,
  &\min\!\left(2z\,|V_{XX}(k)|,\,1\right)\\
  &\times\log\!\left[1-\sin^2\tilde{\alpha}\,\sin^2\Delta_k\right]
\end{split}
\label{eq:Bn_app}
\end{equation}
Eq.~(\ref{eq:Zn_scaling_app}) naturally separates the stationary and dynamical contributions to the charged moments. In particular, $A_2(\tilde{\alpha})$ accounts for the contribution of the initial post-quench state, whereas $B_2(\tilde{\alpha},z)$ describes the dynamical contribution arising from the ballistic propagation of quasiparticles. At $t=0$ and $t \xrightarrow[]{} \infty$ limit, we can recover the expressions of charged moments given in Eq.~(\ref{Z2_ini}) and Eq.~(\ref{eq:Z2_infty_explicit}) respectively.

Having obtained the asymptotic form of the charged moments, we now connect this result to EA. 
Substituting Eq.~(\ref{eq:Zn_scaling_app}) into Eq.~(\ref{Z2_int}) expression leads to,
\begin{equation}
\Delta S_A^{(2)}
=
-\log
\left[
\int_{-\pi}^{\pi}
\frac{d\tilde{\alpha}}{2\pi}
\,
e^{\,
l\left[
A_2(\tilde{\alpha})
+
B_2(\tilde{\alpha},z)
\right]}
\right].
\label{eq:EA_integral}
\end{equation}

To proceed further, we expand the exponential appearing in Eq.~(\ref{eq:EA_integral}) in a Taylor series. Retaining the leading contributions gives

\begin{equation}
\begin{split}
\Delta S_A^{(2)}
=
-\log
\Bigg[
\int_{-\pi}^{\pi}
\frac{d\tilde{\alpha}}{2\pi}
\Big(
1
+
l\big[
A_2(\tilde{\alpha})
+
B_2(\tilde{\alpha},z)
\big]
\\
+\text{higher-order terms}
\Big)
\Bigg]
.
\end{split}
\label{eq:EA_taylor}
\end{equation}
Substituting the explicit expressions for $A_2(\tilde{\alpha})$ and $B_2(\tilde{\alpha},z)$ into Eq.~(\ref{eq:EA_taylor}), we obtain

\begin{equation}
\begin{split}
\Delta S_A^{(2)}
&=
-\log
\Bigg[
1
+
l
\int_{-\pi}^{\pi}
\frac{d\tilde{\alpha}}{2\pi}
\int_{-\pi}^{\pi}
\frac{dk}{4\pi}
\\
&\Big(
1-\min\!\left(2 z |V_{XX}(k)|,1\right)
\Big)
\log\!\Big(
1-\sin^2\tilde{\alpha}\,
\sin^2\Delta_k
\Big)
\Bigg].
\end{split}
\label{eq:EA_after_substitution}
\end{equation}

At this stage, we make use of the standard integral identity

\begin{equation}
\int_{-\pi}^{\pi}
d\tilde{\alpha}\,
\log\!\left(
1-a^2\sin^2\tilde{\alpha}
\right)
=
4\pi
\log\!\left(
\frac{1+\sqrt{1-a^2}}{2}
\right),
\label{eq:standard_integral}
\end{equation}
together with the expansion $\log(1+x)\simeq x$ for $x\rightarrow0$. Consequently, Eq.~(\ref{eq:EA_after_substitution}) reduces to
\begin{align}
\Delta S_A^{(2)}
={}&-l
\int_{-\pi}^{\pi}\frac{d\tilde{\alpha}}{2\pi}
\int_{-\pi}^{\pi}\frac{dk}{4\pi}
\nonumber\\
&\times
\left[1-\min\!\left(2z|V_{XX}(k)|,1\right)\right]
\nonumber\\
&\times
\log\!\left(
1-\sin^2\tilde{\alpha}\,\sin^2\Delta_k
\right).
\label{eq:EA_before_alpha_integral}
\end{align}

Performing the $\tilde{\alpha}$ integration using Eq.~(\ref{eq:standard_integral}), we obtain

\begin{equation}
\begin{split}
\Delta S_A^{(2)}
=
-l
\int_{-\pi}^{\pi}
\frac{dk}{4\pi}
\Big(
1-\min\!\left(2z|V_{XX}(k)|,1\right)
\Big)
\\
\times
\frac{4\pi}{2\pi}
\log\!\left(
\frac{1+\sqrt{1-\sin^2\Delta_k}}{2}
\right),
\end{split}
\label{eq:EA_after_alpha}
\end{equation}
which can be written as
\begin{equation}
\begin{split}
\Delta S_A^{(2)}
=
-2l
\int_{-\pi}^{\pi}
\frac{dk}{4\pi}
\Big(
1-\min\!\left(2z |V_{XX}(k)|,1\right)
\Big)
\\
\times
\log\!\left(
\frac{1+\sqrt{1-\sin^2\Delta_k}}{2}
\right).
\end{split}
\label{eq:EA_final_quasiparticle}
\end{equation}
$\sin^2(\Delta_k)$ in general depends on interaction range parameter $\alpha$ and initial state parameter $\bar{\lambda}$ which is $\theta$ for tilted ferromagnetic states and ${h,\gamma}$ for ground state. We can write Eq.~(\ref{eq:EA_final_quasiparticle}) in a compact form as given by,
\begin{equation}
\Delta S_A^{(2)}
=2lb(z,\alpha,\bar{\lambda})
\label{eq:EA_f}
\end{equation}
where
\begin{equation}
b(z,\alpha,\bar{\lambda}) = \int_{-\pi}^{\pi}\frac{dk}{4\pi}\, x_k(z)~\mathcal{G}_k(\alpha,\bar{\lambda}),
\end{equation}
with
\begin{equation}
\begin{split}
\mathcal{G}_k(\alpha,\bar{\lambda})&=-\ln\left(\frac{1 + \sqrt{1 - \sin^2 \Delta_k}}{2}\right)\\
x_k(z)&=\Big(1-\text{min}(2z|V_{XX}(k)|,1\Big)
\end{split}
\end{equation}
This derivation establishes the connection between the charged moments and the quasiparticle description of the post-quench dynamics. Expressing EA in terms of the asymptotic charged moments provides a microscopic understanding of the late-time behavior underlying the QME.

\subsection{Analytic Expression of $\Delta S^{(2)}_A(t)$ at late times} 
\label{EA_infty}
 Equation (\ref{eq:EA_final_quasiparticle}) has a non-zero value when $|V_{XX}(k)|<\frac{1}{2z}$, and it can be represented as
\begin{equation}
\begin{split}
\Delta S_A^{(2)}(t)
=
-2l
\int_{D}
\frac{dk}{4\pi}
\Big(
1-2z|V_{XX}(k)|
\Big)
\\
\times
\log\!\left(
\frac{1+\sqrt{1-\sin^2\Delta_k}}{2}
\right),
\end{split}
\label{eq:modified EA_final_quasiparticle}
\end{equation}
where $D=\{k:|V_{XX}(k)|<\frac{1}{2z}\}$. {\color{black}At $t \rightarrow \infty$ limit, $\frac{1}{2z}$ goes to zero, hence the domain of integration, $D$ shrinks to a small neighbourhood around $k=k_0$ i.e. $V_{XX}(k_0)=0$}. In this domain, we can approximate $\log\left(
\frac{1+\sqrt{1-\sin^2\Delta_k}}{2}
\right) \approx -\frac{\sin^2{\Delta_k}}{4}$. The analytical expression for EA transforms into
\begin{equation}
    \Delta S^{(2)}_A(t)=l\int_D \frac{dk}{8\pi} \sin^2{\Delta_k}(1-2z|V_{XX}(k)|)
    \label{eq:asymtotic EA_final_quasiparticle}
\end{equation}
In this equation, the initial state contribution comes from the Bogoliubov angle, $\sin^2(\Delta_k)$, which depends on the interaction strength parameter $\alpha$ for the ground state  (Eq.~(\ref{GS})) of the Hamiltonian in Eq.~(\ref{Hamiltonian}), while it is independent of $\alpha$ for the tilted ferromagnetic initial state (Eq.~(\ref{tilted})).
For the tilted ferromagnetic initial states, the quantity $\sin^2\Delta_k$ has the form
\begin{equation}
\sin^2\Delta_k =
\frac{\sin^4\theta\,\sin^2 k}
{\left(1+\cos^2\theta-2\cos\theta\cos k\right)^2},
\label{eq:sin2Delta_tilted}
\end{equation}
and for the ground states
\begin{equation}
\sin^2\Delta_k =
\frac{\gamma^2 R_k^2}
{(h-F_k)^2+\gamma^2 R_k^2},
\label{eq:sin2Delta_ground}
\end{equation}
where
$F_k=\sum_{r=1}^{\infty}r^{-\alpha}\cos(kr)$
and
$R_k=\sum_{r=1}^{\infty}r^{-\alpha}\sin(kr)$.
\par For $0<\alpha<2$, we keep first order contributions in $\sin{\Delta_k}$ and $|V_{XX}(k)|$  at $u=k\pm\pi$, and write Eq.~(\ref{eq:asymtotic EA_final_quasiparticle}) as, 
\begin{equation}
\begin{split}
     \Delta S^{(2)}_A(t)\approx2l\int^{k_\pi}_0 \frac{du}{8\pi} c_{\pm \pi}^2u^2(1-2z|V'_{XX}(\pm\pi)|u).
    \label{eq:asymtotic EA_final_quasiparticle_transformed}
\end{split}
\end{equation}
Here, $|V'_{XX}(\pm\pi)|=|\eta(\alpha-2)|$ and $k_{\pi}=\frac{1}{2z|\eta(\alpha-2)|}$. The integration can be computed exactly,
\begin{equation}
    \Delta S^{(2)}_A(t)\approx\frac{1}{384\pi} \frac{ l^4c_{\pm \pi}^2}{t^3(\eta(\alpha-2))^3},
    \label{asymtotic_EA_longrange}
\end{equation}
where for the tilted ferromagnetic initial state,
\begin{equation}
    c_{\pm\pi}^2 = \tan^4\left(\frac{\theta}{2}\right),
    \label{c_pi_tilted}
\end{equation}
and for the ground state,
\begin{equation}
    c_{\pm\pi}^2 =
    \frac{\gamma^2 \eta^2(\alpha-1)}
    {\left[h+\eta(\alpha)\right]^2}.
    \label{c_pi_GS}
\end{equation}

For $2<\alpha<3$, $V_{XX}(k=0)=0$, but the derivative of $V_{XX}(k)$ diverges at $k=0$.
There, we perform Polylogarithm expansion of $V_{XX}= \text{Im}(\text{Li}_{\alpha-1}(e^{ik}))$ around $k=0$. The leading order term is given by $|V_{XX}(0\pm u)|\approx |\Gamma(2-\alpha)|\text{sin}(\frac{\pi(\alpha-2)}{2})|u|^{\alpha-2}$.  The contribution from slow modes around $k=\pm\pi$ is identical to Eq.~(\ref{asymtotic_EA_longrange}). With this, we can write Eq.~(\ref{eq:asymtotic EA_final_quasiparticle}) as, 
\begin{equation}
\begin{split}
     \Delta S^{(2)}_A(t) &\approx 2l\int^{k_\pi}_0 \frac{du}{8\pi} c_{\pm \pi}^2u^2(1-2zV'_{XX}(\pm\pi)|u) \\
     &+2l\int^{k_0}_0 \frac{du}{8\pi} c_{0}^2u^2(1-2z|V_{XX}(0\pm u)|),
\end{split}
\end{equation}
with $k_0=\big(2z \Gamma(2-\alpha)|\sin(\frac{\pi(\alpha-2)}{2}|\big)^{\frac{1}{2-\alpha}}$.Performing  the integration exactly, we obtain EA in the asymptotic limit as      
\begin{equation}
\begin{split}
\Delta S^{(2)}_A(t) &\approx
\frac{(\alpha-2)\,c_0^2l^{\frac{\alpha+1}{\alpha-2}}}
     {12\pi(\alpha+1)\Big[\big|2t\Gamma(2-\alpha)\,\sin\!\big(\tfrac{\pi(\alpha-2)}{2}\big)\big|\Big]^{\frac{3}{\alpha-2}}}\\[4pt]
&\quad+\frac{c_{\pm\pi}^2}{384\pi\,(\eta(\alpha-2))^{3}}\frac{l^4}{t^{3}},
\end{split}
\label{asymmetry 2to3}
\end{equation}
where for the tilted ferromagnetic initial state (Eq.~\eqref{tilted}),
\begin{equation}
    c_0^2 = \cot^4\left(\frac{\theta}{2}\right),
    \label{c_0_tilted}
\end{equation}
and for the ground state of Eq.~\eqref{GS},
\begin{equation}
    c_0^2 =
    \frac{\gamma^2 \zeta^2(\alpha-1)}
    {\left[h-\zeta(\alpha)\right]^2}.
    \label{c_0_GS}
\end{equation}
\par In Eq.~(\ref{asymmetry 2to3}), the contribution from $k=0$ is subleading with $t^{-3/(\alpha-2)}$ compared to $k=\pm\pi$ contribution of order $t^{-3}$. Therefore, at large times, the contribution from $k=\pm\pi$ dominates and in that regime we express the asymptotic expression for EA as given in Eq.~(\ref{asymtotic_EA_longrange}).

For $\alpha=3$, the dominating term of expansion of $V_{XX}(k)$ around $k=0$ is $|V_{XX}(0\pm u)|\approx |u|(1+\text{ln}\big(\frac{1}{|u|})\big)$
with $|u|\ll1$. Using this, we calculate the contribution of slow modes around $k=0$ in late-time EA. The contribution from slow modes around $k=\pm\pi$ is similar to $\alpha<3$ regime:
\begin{equation}
\begin{split}
    \Delta S^{(2)}_A(t)&\approx\frac{c_0^2}{384\pi}\frac{\ln(\frac{1}{k_e})+\frac{1}{4}}{\Big(1+\ln(\frac{1}{k_e})\Big)^4}\times\frac{l^4}{t^3}\\
    &+\frac{c_{\pm\pi}^2}{384\pi\,(\eta(\alpha-2))^{3}}\,\frac{l^4}{t^{3}},
\end{split}
\label{tilted_ferro_3}
\end{equation}
where $k_e$ is the positive root of the equation,
\begin{equation}
    \frac{1}{2z k_e}-1=\ln\Big(\frac{1}{k_e}\Big).
    \label{root_alpha3}
\end{equation}
Now, for $\alpha> 3$,
the group velocity, $|V_{XX}(k)|$, around $k=0$ is given by,  $|V_{XX}(0\pm u)|\approx\zeta(\alpha-2)|u|$, with $|u|\ll1$. The asymptotic expression for EA in this regime turns out as,
\begin{equation}
    \Delta S^{(2)}_A(t)\approx\frac{c_{\pm \pi}^2}{384\pi\,(\eta(\alpha-2))^{3}}\,\frac{l^4}{t^{3}}+\frac{c_0^2}{384\pi\,(\zeta(\alpha-2))^{3}}\,\frac{l^4}{t^{3}}.
    \label{tilted_ferromagnetic_geq4}
\end{equation}
Note that the critical magnetic field $h_c=\zeta(\alpha)$
limits the validity of Eqs.~\eqref{asymmetry 2to3},
\eqref{tilted_ferro_3}, and \eqref{tilted_ferromagnetic_geq4},which apply only away from criticality.

\newpage

}

\bibliography{References}

@article{SciPostPhys.15.3.089,
	title = {Lack of symmetry restoration after a quantum quench: An entanglement asymmetry study},
	pages = {089},
	author = {Ares, Filiberto and Murciano, Sara and Vernier, Eric and Calabrese, Pasquale},
	journal = {SciPost Phys.},
	volume = {15},
	year = {2023},
	publisher = {SciPost},
	doi = {10.21468/SciPostPhys.15.3.089},
	url = {https://scipost.org/10.21468/SciPostPhys.15.3.089}
}

@article{Calabrese_2005,
doi = {10.1088/1742-5468/2005/04/P04010},
url = {https://doi.org/10.1088/1742-5468/2005/04/P04010},
year = {2005},
month = {apr},
publisher = {},
volume = {2005},
number = {04},
pages = {P04010},
author = {Calabrese, Pasquale and Cardy, John},
title = {Evolution of entanglement entropy in one-dimensional systems},
journal = {Journal of Statistical Mechanics: Theory and Experiment}
}

@article{Alba-Calabrese_2017,
author = {Vincenzo Alba  and Pasquale Calabrese },
title = {Entanglement and thermodynamics after a quantum quench in integrable systems},
journal = {Proceedings of the National Academy of Sciences},
volume = {114},
number = {30},
pages = {7947-7951},
year = {2017},
doi = {10.1073/pnas.1703516114},
URL = {https://www.pnas.org/doi/abs/10.1073/pnas.1703516114},
}

@article{Murciano2024,
  title = {Entanglement asymmetry and quantum Mpemba effect in the XY spin chain},
  volume = {2024},
  ISSN = {1742-5468},
  url = {http://dx.doi.org/10.1088/1742-5468/ad17b4},
  DOI = {10.1088/1742-5468/ad17b4},
  number = {1},
  journal = {Journal of Statistical Mechanics: Theory and Experiment},
  publisher = {IOP Publishing},
  author = {Murciano,  Sara and Ares,  Filiberto and Klich,  Israel and Calabrese,  Pasquale},
  year = {2024},
  month = Jan,
  pages = {013103}
}

@article{Auerbach1995,
  title = {Supercooling and the Mpemba effect: When hot water freezes quicker than cold},
  volume = {63},
  ISSN = {1943-2909},
  url = {http://dx.doi.org/10.1119/1.18059},
  DOI = {10.1119/1.18059},
  number = {10},
  journal = {American Journal of Physics},
  publisher = {American Association of Physics Teachers (AAPT)},
  author = {Auerbach,  David},
  year = {1995},
  month = Oct,
  pages = {882–885}
}

@article{Jeng2006,
  title = {The Mpemba effect: When can hot water freeze faster than cold?},
  volume = {74},
  ISSN = {1943-2909},
  url = {http://dx.doi.org/10.1119/1.2186331},
  DOI = {10.1119/1.2186331},
  number = {6},
  journal = {American Journal of Physics},
  publisher = {American Association of Physics Teachers (AAPT)},
  author = {Jeng,  Monwhea},
  year = {2006},
  month = June,
  pages = {514–522}
}

@article{Burridge2016,
  title = {Questioning the Mpemba effect: hot water does not cool more quickly than cold},
  volume = {6},
  ISSN = {2045-2322},
  url = {http://dx.doi.org/10.1038/srep37665},
  
  number = {1},
  journal = {Scientific Reports},
  publisher = {Springer Science and Business Media LLC},
  author = {Burridge,  Henry C. and Linden,  Paul F.},
  year = {2016},
  month = Nov 
}

@article{Kumar2020,
  title = {Exponentially faster cooling in a colloidal system},
  volume = {584},
  ISSN = {1476-4687},
  url = {http://dx.doi.org/10.1038/s41586-020-2560-x},
  DOI = {10.1038/s41586-020-2560-x},
  number = {7819},
  journal = {Nature},
  publisher = {Springer Science and Business Media LLC},
  author = {Kumar,  Avinash and Bechhoefer,  John},
  year = {2020},
  month = Aug,
  pages = {64–68}
}

@article{Kumar2022,
  title = {Anomalous heating in a colloidal system},
  author = {Kumar,  Avinash and Chétrite,  Raphaël and Bechhoefer,  John},
  volume = {119},
  ISSN = {1091-6490},
  url = {http://dx.doi.org/10.1073/pnas.2118484119},
 
  number = {5},
  journal = {Proceedings of the National Academy of Sciences},
  publisher = {National Academy of Sciences},
  
  year = {2022},
  month = Jan 
}

@article{PhysRevLett.119.148001,
  title = {When the Hotter Cools More Quickly: Mpemba Effect in Granular Fluids},
  author = {Lasanta, Antonio and V. Reyes, Francisco and Prados, Antonio and Santos, Andr\'es},
  journal = {Phys. Rev. Lett.},
  volume = {119},
  issue = {14},
  pages = {148001},
  numpages = {6},
  year = {2017},
  month = {Oct},
  publisher = {American Physical Society},
  doi = {10.1103/PhysRevLett.119.148001},
  url = {https://link.aps.org/doi/10.1103/PhysRevLett.119.148001}
}

@article{Lu2017,
  title = {Nonequilibrium thermodynamics of the Markovian Mpemba effect and its inverse},
  volume = {114},
  ISSN = {1091-6490},
  url = {http://dx.doi.org/10.1073/pnas.1701264114},
  DOI = {10.1073/pnas.1701264114},
  number = {20},
  journal = {Proceedings of the National Academy of Sciences},
  publisher = {National Academy of Sciences},
  author = {Lu,  Zhiyue and Raz,  Oren},
  year = {2017},
  month = May,
  pages = {5083–5088}
}

@article{PhysRevX.9.021060,
  title = {Mpemba Index and Anomalous Relaxation},
  author = {Klich, Israel and Raz, Oren and Hirschberg, Ori and Vucelja, Marija},
  journal = {Phys. Rev. X},
  volume = {9},
  issue = {2},
  pages = {021060},
  numpages = {17},
  year = {2019},
  month = {Jun},
  publisher = {American Physical Society},
  doi = {10.1103/PhysRevX.9.021060},
  url = {https://link.aps.org/doi/10.1103/PhysRevX.9.021060}
}

@article{PhysRevE.102.012906,
  title = {Mpemba effect in driven granular Maxwell gases},
  author = {Biswas, Apurba and Prasad, V. V. and Raz, O. and Rajesh, R.},
  journal = {Phys. Rev. E},
  volume = {102},
  issue = {1},
  pages = {012906},
  numpages = {14},
  year = {2020},
  month = {Jul},
  publisher = {American Physical Society},
  doi = {10.1103/PhysRevE.102.012906},
  url = {https://link.aps.org/doi/10.1103/PhysRevE.102.012906}
}

@article{Biswas2021,
  title = {Mpemba effect in an anisotropically driven granular gas},
  volume = {136},
  ISSN = {1286-4854},
  url = {http://dx.doi.org/10.1209/0295-5075/ac2d54},
  DOI = {10.1209/0295-5075/ac2d54},
  number = {4},
  journal = {Europhysics Letters},
  publisher = {IOP Publishing},
  author = {Biswas,  Apurba and Prasad,  V. V. and Rajesh,  R.},
  year = {2021},
  month = Nov,
  pages = {46001}
}

@article{Holtzman2022,
  title   = {Landau theory for the {Mpemba} effect through phase transitions},
  author  = {Holtzman, Roi and Raz, Oren},
  journal = {Commun. Phys.},
  volume  = {5},
  number  = {1},
  pages   = {280},
  year    = {2022},
  doi     = {10.1038/s42005-022-01063-2},
  url     = {https://doi.org/10.1038/s42005-022-01063-2}
}

@article{Santos2020,
  title = {Mpemba effect in molecular gases under nonlinear drag},
  volume = {32},
  ISSN = {1089-7666},
  url = {http://dx.doi.org/10.1063/5.0016243},
 
  number = {7},
  journal = {Physics of Fluids},
  publisher = {AIP Publishing},
  author = {Santos,  Andrés and Prados,  Antonio},
  year = {2020},
  month = July 
}

@article{Ares2025,
  title = {The quantum Mpemba effects},
  volume = {7},
  ISSN = {2522-5820},
  url = {http://dx.doi.org/10.1038/s42254-025-00838-0},
  DOI = {10.1038/s42254-025-00838-0},
  number = {8},
  journal = {Nature Reviews Physics},
  publisher = {Springer Science and Business Media LLC},
  author = {Ares,  Filiberto and Calabrese,  Pasquale and Murciano,  Sara},
  year = {2025},
  month = July,
  pages = {451–460}
}

@article{PhysRevLett.131.080402,
  title = {Quantum Mpemba Effect in a Quantum Dot with Reservoirs},
  author = {Chatterjee, Amit Kumar and Takada, Satoshi and Hayakawa, Hisao},
  journal = {Phys. Rev. Lett.},
  volume = {131},
  issue = {8},
  pages = {080402},
  numpages = {6},
  year = {2023},
  month = {Aug},
  publisher = {American Physical Society},
  doi = {10.1103/PhysRevLett.131.080402},
  url = {https://link.aps.org/doi/10.1103/PhysRevLett.131.080402}
}

@article{PhysRevA.110.022213,
  title = {Multiple quantum Mpemba effect: Exceptional points and oscillations},
  author = {Chatterjee, Amit Kumar and Takada, Satoshi and Hayakawa, Hisao},
  journal = {Phys. Rev. A},
  volume = {110},
  issue = {2},
  pages = {022213},
  numpages = {21},
  year = {2024},
  month = {Aug},
  publisher = {American Physical Society},
  doi = {10.1103/PhysRevA.110.022213},
  url = {https://link.aps.org/doi/10.1103/PhysRevA.110.022213}
}

@article{PhysRevResearch.6.033330,
  title = {Mpemba effects in nonequilibrium open quantum systems},
  author = {Wang, Xuanhua and Wang, Jin},
  journal = {Phys. Rev. Res.},
  volume = {6},
  issue = {3},
  pages = {033330},
  numpages = {7},
  year = {2024},
  month = {Sep},
  publisher = {American Physical Society},
  doi = {10.1103/PhysRevResearch.6.033330},
  url = {https://link.aps.org/doi/10.1103/PhysRevResearch.6.033330}
}

@article{PhysRevA.110.042218,
  title = {Speeding up quantum heat engines by the Mpemba effect},
  author = {Liu, Dehua and Yuan, Jiehong and Ruan, Huilin and Xu, Yang and Luo, Shaolin and He, Jizhou and He, Xian and Ma, Yongli and Wang, Jianhui},
  journal = {Phys. Rev. A},
  volume = {110},
  issue = {4},
  pages = {042218},
  numpages = {12},
  year = {2024},
  month = {Oct},
  publisher = {American Physical Society},
  doi = {10.1103/PhysRevA.110.042218},
  url = {https://link.aps.org/doi/10.1103/PhysRevA.110.042218}
}

@article{8fp4-15mq,
  title = {Mpemba effect in self-contained quantum refrigerators: Accelerated cooling},
  author = {Mondal, Sayan and Sen, Ujjwal},
  journal = {Phys. Rev. A},
  volume = {113},
  issue = {2},
  pages = {022211},
  numpages = {12},
  year = {2026},
  month = {Feb},
  publisher = {American Physical Society},
  doi = {10.1103/8fp4-15mq},
  url = {https://link.aps.org/doi/10.1103/8fp4-15mq}
}

@article{das2025rolereversalquantummpemba,
  title={Role reversal in quantum mpemba effect},
  author={Das, Arunabha and Chaki, Paranjoy and Ghosh, Priya and Sen, Ujjwal},
  journal={arXiv:2512.24839},
  year={2025},
  url={https://arxiv.org/abs/2512.24839}}

@article{bagui2025detectionmpembaeffectgood,
  title={Detection of Mpemba effect through good observables in open quantum systems},
  author={Bagui, Pitambar and Chatterjee, Arijit and Agarwalla, Bijay Kumar},
  journal={arXiv:2512.02709},
  year={2025},
 url={https://arxiv.org/abs/2512.02709},
}

@article{Chalas_2024,
doi = {10.1088/1742-5468/ad769c},
url = {https://doi.org/10.1088/1742-5468/ad769c},
year = {2024},
month = {nov},
publisher = {IOP Publishing},
volume = {2024},
number = {10},
pages = {103101},
author = {Chalas, Konstantinos and Ares, Filiberto and Rylands, Colin and Calabrese, Pasquale},
title = {Multiple crossings during dynamical symmetry restoration and implications for the quantum Mpemba effect},
journal = {Journal of Statistical Mechanics: Theory and Experiment}
}

@article{Lastres_2025,
doi = {10.1088/1742-5468/ada497},
url = {https://doi.org/10.1088/1742-5468/ada497},
year = {2025},
month = {jan},
publisher = {IOP Publishing},
volume = {2025},
number = {1},
pages = {013107},
author = {Lastres, Marco and Murciano, Sara and Ares, Filiberto and Calabrese, Pasquale},
title = {Entanglement asymmetry in the critical XXZ spin chain},
journal = {Journal of Statistical Mechanics: Theory and Experiment}
}

@article{PhysRevB.111.104312,
  title = {Quantum Mpemba effect in free-fermionic mixed states},
  author = {Ares, Filiberto and Vitale, Vittorio and Murciano, Sara},
  journal = {Phys. Rev. B},
  volume = {111},
  issue = {10},
  pages = {104312},
  numpages = {17},
  year = {2025},
  month = {Mar},
  publisher = {American Physical Society},
  doi = {10.1103/PhysRevB.111.104312},
  url = {https://link.aps.org/doi/10.1103/PhysRevB.111.104312}
}

@article{PhysRevB.110.085126,
  title = {Entanglement asymmetry and quantum Mpemba effect in two-dimensional free-fermion systems},
  author = {Yamashika, Shion and Ares, Filiberto and Calabrese, Pasquale},
  journal = {Phys. Rev. B},
  volume = {110},
  issue = {8},
  pages = {085126},
  numpages = {16},
  year = {2024},
  month = {Aug},
  publisher = {American Physical Society},
  doi = {10.1103/PhysRevB.110.085126},
  url = {https://link.aps.org/doi/10.1103/PhysRevB.110.085126}
}

@article{PhysRevA.111.043304,
  title = {Quenching from superfluid to free bosons in two dimensions: Entanglement, symmetries, and the quantum Mpemba effect},
  author = {Yamashika, Shion and Calabrese, Pasquale and Ares, Filiberto},
  journal = {Phys. Rev. A},
  volume = {111},
  issue = {4},
  pages = {043304},
  numpages = {28},
  year = {2025},
  month = {Apr},
  publisher = {American Physical Society},
  doi = {10.1103/PhysRevA.111.043304},
  url = {https://link.aps.org/doi/10.1103/PhysRevA.111.043304}
}

@article{l4xx-plmv,
  title = {Quantum Mpemba effect in long-ranged U(1)-symmetric random circuits},
  author = {Li, Han-Ze and Lee, Ching Hua and Liu, Shuo and Zhang, Shi-Xin and Zhong, Jian-Xin},
  journal = {Phys. Rev. B},
  volume = {113},
  issue = {13},
  pages = {134310},
  numpages = {15},
  year = {2026},
  month = {Apr},
  publisher = {American Physical Society},
  doi = {10.1103/l4xx-plmv},
  url = {https://link.aps.org/doi/10.1103/l4xx-plmv}
}

@article{1td3-2vwf,
  title = {Quantum Mpemba effect without global symmetries},
  author = {Bhore, Tanmay and Su, Lei and Martin, Ivar and Clerk, Aashish A. and Papi\ifmmode \acute{c}\else \'{c}\fi{}, Zlatko},
  journal = {Phys. Rev. B},
  volume = {112},
  issue = {12},
  pages = {L121109},
  numpages = {6},
  year = {2025},
  month = {Sep},
  publisher = {American Physical Society},
  doi = {10.1103/1td3-2vwf},
  url = {https://link.aps.org/doi/10.1103/1td3-2vwf}
}

@article{fujimura2026entanglementasymmetryquantummpemba,
  title   = {Entanglement asymmetry and quantum {Mpemba} effect for non-{Abelian} global symmetry},
  author  = {Fujimura, Harunobu and Shimamori, Soichiro},
  journal = {J. High Energy Phys.},
  volume  = {03},
  pages   = {244},
  year    = {2026},
  doi     = {10.1007/JHEP03(2026)244},
  url     = {https://doi.org/10.1007/JHEP03(2026)244}
}

@article{yu2025quantum,
  title   = {Quantum {Mpemba} effects from symmetry perspectives},
  author  = {Yu, Hui and Liu, Shuo and Zhang, Shi-Xin},
  journal = {AAPPS Bull.},
  volume  = {35},
  number  = {1},
  pages   = {17},
  year    = {2025},
  doi     = {10.1007/s43673-025-00157-7},
  url     = {https://doi.org/10.1007/s43673-025-00157-7}
}

@article{mpemba1969cool,
  title   = {Cool?},
  author  = {Mpemba, Erasto B. and Osborne, Denis G.},
  journal = {Phys. Educ.},
  volume  = {4},
  number  = {3},
  pages   = {172--175},
  year    = {1969},
  doi     = {10.1088/0031-9120/4/3/312},
  url     = {https://doi.org/10.1088/0031-9120/4/3/312}
}

@article{alishahiha2025eigenstatethermalizationhypothesisshort,
  title={Eigenstate thermalization hypothesis: A Primer},
  author={Alishahiha, Mohsen and Vasli, Mohammad Javad},
  journal={arXiv:2501.07243},
  url = {https://arxiv.org/abs/2501.07243},
  year={2025}
}

@article{Deutsch_2018,
doi = {10.1088/1361-6633/aac9f1},
url = {https://doi.org/10.1088/1361-6633/aac9f1},
year = {2018},
month = {jul},
publisher = {IOP Publishing},
volume = {81},
number = {8},
pages = {082001},
author = {Deutsch, Joshua M},
title = {Eigenstate thermalization hypothesis},
journal = {Reports on Progress in Physics}
}

@article{PhysRevLett.127.060401,
  title = {Exponentially Accelerated Approach to Stationarity in Markovian Open Quantum Systems through the Mpemba Effect},
  author = {Carollo, Federico and Lasanta, Antonio and Lesanovsky, Igor},
  journal = {Phys. Rev. Lett.},
  volume = {127},
  issue = {6},
  pages = {060401},
  numpages = {6},
  year = {2021},
  month = {Aug},
  publisher = {American Physical Society},
  doi = {10.1103/PhysRevLett.127.060401},
  url = {https://link.aps.org/doi/10.1103/PhysRevLett.127.060401}
}

@article{PhysRevLett.133.140404,
  title = {Thermodynamics of the Quantum Mpemba Effect},
  author = {Moroder, Mattia and Culhane, Ois\'{\i}n and Zawadzki, Krissia and Goold, John},
  journal = {Phys. Rev. Lett.},
  volume = {133},
  issue = {14},
  pages = {140404},
  numpages = {6},
  year = {2024},
  month = {Oct},
  publisher = {American Physical Society},
  doi = {10.1103/PhysRevLett.133.140404},
  url = {https://link.aps.org/doi/10.1103/PhysRevLett.133.140404}
}

@article{PhysRevLett.133.136302,
  title = {Mpemba Effects in Open Nonequilibrium Quantum Systems},
  author = {Nava, Andrea and Egger, Reinhold},
  journal = {Phys. Rev. Lett.},
  volume = {133},
  issue = {13},
  pages = {136302},
  numpages = {7},
  year = {2024},
  month = {Sep},
  publisher = {American Physical Society},
  doi = {10.1103/PhysRevLett.133.136302},
  url = {https://link.aps.org/doi/10.1103/PhysRevLett.133.136302}
}

@article{PhysRevLett.133.010401,
  title = {Microscopic Origin of the Quantum Mpemba Effect in Integrable Systems},
  author = {Rylands, Colin and Klobas, Katja and Ares, Filiberto and Calabrese, Pasquale and Murciano, Sara and Bertini, Bruno},
  journal = {Phys. Rev. Lett.},
  volume = {133},
  issue = {1},
  pages = {010401},
  numpages = {6},
  year = {2024},
  month = {Jul},
  publisher = {American Physical Society},
  doi = {10.1103/PhysRevLett.133.010401},
  url = {https://link.aps.org/doi/10.1103/PhysRevLett.133.010401}
}

@article{Vodola_2016,
doi = {10.1088/1367-2630/18/1/015001},
url = {https://doi.org/10.1088/1367-2630/18/1/015001},
year = {2015},
month = {dec},
publisher = {IOP Publishing},
volume = {18},
number = {1},
pages = {015001},
author = {Vodola, Davide and Lepori, Luca and Ercolessi, Elisa and Pupillo, Guido},
title = {Long-range Ising and Kitaev models: phases, correlations and edge modes},
journal = {New Journal of Physics}
}

@article{ares2023entanglement,
  title   = {Entanglement asymmetry as a probe of symmetry breaking},
  author  = {Ares, Filiberto and Murciano, Sara and Calabrese, Pasquale},
  journal = {Nat. Commun.},
  volume  = {14},
  number  = {1},
  pages   = {2036},
  year    = {2023},
  doi     = {10.1038/s41467-023-37747-8},
  url     = {https://doi.org/10.1038/s41467-023-37747-8}
}

@article{schnepper2025experimentalobservationapplicationgenuine,
  title={Experimental observation and application of the genuine quantum Mpemba effect},
  author={Schnepper, Bruno P and de Oliveira, Jefferson LD and Av{\'o}, Yan AC and Vieira, Carlos HS and Zawadzki, Krissia and Serra, Roberto M},
  journal={arXiv:2511.14552},
  url={https://arxiv.org/abs/2511.14552},
  year={2025}
}

@article{chatterjee2025directexperimentalobservationquantum,
  title={Direct experimental observation of quantum Mpemba effect without bath engineering},
  author={Chatterjee, Arijit and Khan, Sakil and Jain, Sachin and Mahesh, TS},
  journal={arXiv:2509.13451},
  url = {https://arxiv.org/abs/2509.13451},
  year={2025}
}

@article{zhang2025observation,
  title   = {Observation of quantum strong {Mpemba} effect},
  author  = {Zhang, Jie and Xia, Gang and Wu, Chun-Wang and Chen, Ting and Zhang, Qian and Xie, Yi and Su, Wen-Bo and Wu, Wei and Qiu, Cheng-Wei and Chen, Ping-Xing and Li, Weibin and Jing, Hui and Zhou, Yan-Li},
  journal = {Nat. Commun.},
  volume  = {16},
  number  = {1},
  pages   = {301},
  year    = {2025},
  doi     = {10.1038/s41467-024-54303-0},
  url     = {https://doi.org/10.1038/s41467-024-54303-0}
}

@article{RevModPhys.95.035002,
  title = {Long-range interacting quantum systems},
  author = {Defenu, Nicol\`o and Donner, Tobias and Macr\`{\i}, Tommaso and Pagano, Guido and Ruffo, Stefano and Trombettoni, Andrea},
  journal = {Rev. Mod. Phys.},
  volume = {95},
  issue = {3},
  pages = {035002},
  numpages = {70},
  year = {2023},
  month = {Aug},
  publisher = {American Physical Society},
  doi = {10.1103/RevModPhys.95.035002},
  url = {https://link.aps.org/doi/10.1103/RevModPhys.95.035002}
}

@article{RevModPhys.82.2313,
  title = {Quantum information with Rydberg atoms},
  author = {Saffman, M. and Walker, T. G. and M\o{}lmer, K.},
  journal = {Rev. Mod. Phys.},
  volume = {82},
  issue = {3},
  pages = {2313--2363},
  numpages = {0},
  year = {2010},
  month = {Aug},
  publisher = {American Physical Society},
  doi = {10.1103/RevModPhys.82.2313},
  url = {https://link.aps.org/doi/10.1103/RevModPhys.82.2313}
}

@article{Lahaye_2009,
doi = {10.1088/0034-4885/72/12/126401},
url = {https://doi.org/10.1088/0034-4885/72/12/126401},
year = {2009},
month = {nov},
publisher = {},
volume = {72},
number = {12},
pages = {126401},
author = {Lahaye, T and Menotti, C and Santos, L and Lewenstein, M and Pfau, T},
title = {The physics of dipolar bosonic quantum gases},
journal = {Reports on Progress in Physics}
}

@article{Carr_2009,
doi = {10.1088/1367-2630/11/5/055049},
url = {https://doi.org/10.1088/1367-2630/11/5/055049},
year = {2009},
month = {may},
publisher = {IOP Publishing},
volume = {11},
number = {5},
pages = {055049},
author = {Carr, Lincoln D and DeMille, David and Krems, Roman V and Ye, Jun},
title = {Cold and ultracold molecules: science, technology and
               applications},
journal = {New Journal of Physics}
}

@article{Blatt:2012chk,
    author = "Blatt, R. and Roos, C. F.",
    title = "{Quantum simulations with trapped ions}",
    doi = "10.1038/nphys2252",
    journal = "Nature Phys.",
    volume = "8",
    pages = "277--284",
    year = "2012"
}

@ARTICLE{2008Natur.453.1008B,
       author = {{Blatt}, Rainer and {Wineland}, David},
        title = "{Entangled states of trapped atomic ions}",
      journal = {\nat},
         year = 2008,
        month = jun,
       volume = {453},
       number = {7198},
        pages = {1008-1015},
          doi = {10.1038/nature07125},
       adsurl = {https://ui.adsabs.harvard.edu/abs/2008Natur.453.1008B}
}

@article{RevModPhys.93.025001,
  title = {Programmable quantum simulations of spin systems with trapped ions},
  author = {Monroe, C. and Campbell, W. C. and Duan, L.-M. and Gong, Z.-X. and Gorshkov, A. V. and Hess, P. W. and Islam, R. and Kim, K. and Linke, N. M. and Pagano, G. and Richerme, P. and Senko, C. and Yao, N. Y.},
  journal = {Rev. Mod. Phys.},
  volume = {93},
  issue = {2},
  pages = {025001},
  numpages = {57},
  year = {2021},
  month = {Apr},
  publisher = {American Physical Society},
  doi = {10.1103/RevModPhys.93.025001},
  url = {https://link.aps.org/doi/10.1103/RevModPhys.93.025001}
}

@article{PhysRevLett.133.010402,
  title = {Observing the Quantum Mpemba Effect in Quantum Simulations},
  author = {Joshi, Lata Kh. and Franke, Johannes and Rath, Aniket and Ares, Filiberto and Murciano, Sara and Kranzl, Florian and Blatt, Rainer and Zoller, Peter and Vermersch, Beno\^{\i}t and Calabrese, Pasquale and Roos, Christian F. and Joshi, Manoj K.},
  journal = {Phys. Rev. Lett.},
  volume = {133},
  issue = {1},
  pages = {010402},
  numpages = {7},
  year = {2024},
  month = {Jul},
  publisher = {American Physical Society},
  doi = {10.1103/PhysRevLett.133.010402},
  url = {https://link.aps.org/doi/10.1103/PhysRevLett.133.010402}
}

@article{hallam2025tunablequantummpembaeffect,
  title = {Tunable quantum Mpemba effect in long-range interacting systems},
  author = {Hallam, Andrew and Yusuf, Matthew and Clerk, Aashish A. and Martin, Ivar and Papi{\ifmmode \acute{c}\else \'{c}\fi{}}, Zlatko},
  journal = {Phys. Rev. B},
  volume = {114},
  issue = {12},
  pages = {125135},
  numpages = {11},
  year = {2026},
  month = {Aug},
  publisher = {American Physical Society},
  doi = {10.1103/kvth-my1s},
  url = {https://link.aps.org/doi/10.1103/kvth-my1s}
}

@article{Xu2026,
  author  = {Xu, Yueshan and Fang, Cai-Ping and Chen, Bing-Jie and Wang, Ming-Chuan and others},
  title   = {Observation and Modulation of the Quantum {Mpemba} Effect on a Superconducting Quantum Processor},
  journal = {Phys. Rev. Lett.},
  volume  = {137},
  pages   = {010402},
  year    = {2026},
  doi     = {10.1103/951q-j8kq}
}

@article{AharonyShapira2024,
  title = {Inverse Mpemba Effect Demonstrated on a Single Trapped Ion Qubit},
  author = {Aharony Shapira, Shahaf and Shapira, Yotam and Markov, Jovan and Teza, Gianluca and Akerman, Nitzan and Raz, Oren and Ozeri, Roee},
  journal = {Phys. Rev. Lett.},
  volume = {133},
  issue = {1},
  pages = {010403},
  numpages = {6},
  year = {2024},
  month = {Jul},
  publisher = {American Physical Society},
  doi = {10.1103/PhysRevLett.133.010403},
  url = {https://link.aps.org/doi/10.1103/PhysRevLett.133.010403}
}

@article{Fraxanet2021,
  author   = {Fraxanet, Joana and Bhattacharya, Utso and Grass, Tobias and Rakshit, Debraj and Lewenstein, Maciej and Dauphin, Alexandre},
  title    = {Topological properties of the long-range {K}itaev chain with {A}ubry-{A}ndr{\'e}-{H}arper modulation},
  journal  = {Phys. Rev. Research},
  volume   = {3},
  number   = {1},
  pages    = {013148},
  numpages = {18},
  year     = {2021},
  month    = feb,
  doi      = {10.1103/PhysRevResearch.3.013148}
}

@misc{NIST,
  key          = "{\relax DLMF}",
  title        = "{NIST Digital Library of Mathematical Functions}",
  editor       = {Olver, F. W. J. and Olde Daalhuis, A. B. and Lozier, D. W. and
  Schneider, B. I. and
  Boisvert, R. F. and
  Clark, C. W. and
  Miller, B. R. and
  Saunders, B. V. and
  Cohl, H. S. and
  McClain, M. A.},
  howpublished = {\url{https://dlmf.nist.gov/}},
  note         = {Release 1.2.8, September 15, 2026;
  see Eq. (25.12.12)},url     = {https://dlmf.nist.gov/25.12.E12}
}

@article{Konar2025,
  author  = {Konar, Tanoy Kanti and Lakkaraju, Leela Ganesh Chandra and Sen(De), Aditi},
  title   = {Slow decay rate of correlations induced by long-range extended {Dzyaloshinskii-Moriya} interactions},
  journal = {Phys. Rev. A},
  volume  = {111},
  pages   = {052402},
  year    = {2025},
  month   = may,
  doi     = {10.1103/PhysRevA.111.052402}
}

@article{Tran2020,
  author  = {Tran, Minh C. and Chen, Chi-Fang and Ehrenberg, Adam and Guo, Andrew Y. and
  Deshpande, Abhinav and Hong, Yifan and Gong, Zhe-Xuan and
  Gorshkov, Alexey V. and Lucas, Andrew},
  title   = {Hierarchy of Linear Light Cones with Long-Range Interactions},
  journal = {Phys. Rev. X},
  volume  = {10},
  pages   = {031009},
  year    = {2020},
  doi     = {10.1103/PhysRevX.10.031009}
}

@article{Yamashika2026,
  title = {Quantum Mpemba Effect in Long-Range Spin Systems},
  author = {Yamashika, Shion and Ares, Filiberto},
  journal = {Phys. Rev. Lett.},
  volume = {136},
  issue = {9},
  pages = {090402},
  numpages = {7},
  year = {2026},
  month = {Mar},
  publisher = {American Physical Society},
  doi = {10.1103/52y5-8kl2},
  url = {https://link.aps.org/doi/10.1103/52y5-8kl2}
}

@article{Jiang2018, 
author={Jiang, Zhang and Sung, Kevin J. and Kechedzhi, Kostyantyn and Smelyanskiy, Vadim N. and Boixo, Sergio},
  title={Quantum Algorithms to Simulate Many-Body Physics of Correlated Fermions},
  journal={Phys. Rev. Applied}, volume={9}, pages={044036}, year={2018},
  doi={10.1103/PhysRevApplied.9.044036}}

@article{Kivlichan2018, 
author={Kivlichan, Ian D. and McClean, Jarrod and Wiebe, Nathan and Gidney, Craig and Aspuru-Guzik, Al{\'a}n and Chan, Garnet Kin-Lic and Babbush, Ryan},
  title={Quantum Simulation of Electronic Structure with Linear Depth and Connectivity},
  journal={Phys. Rev. Lett.}, volume={120}, pages={110501}, year={2018},
  doi={10.1103/PhysRevLett.120.110501}}

 \end{document}